\documentclass[lettersize,journal]{IEEEtran}
\IEEEoverridecommandlockouts
\usepackage{cite}
\usepackage{amsmath,amssymb,amsfonts}
\usepackage{algorithmic}
\usepackage{graphicx}
\usepackage{textcomp}
\usepackage{xcolor}
\def\BibTeX{{\rm B\kern-.05em{\sc i\kern-.025em b}\kern-.08em
    T\kern-.1667em\lower.7ex\hbox{E}\kern-.125emX}}
\usepackage{multirow}
\usepackage{algorithm}
\usepackage{array}
\usepackage[caption=false,font=normalsize,labelfont=sf,textfont=sf]{subfig}
\usepackage{textcomp}
\usepackage{stfloats}
\usepackage{url}
\usepackage{xcolor}
\usepackage{array}
\usepackage{verbatim}
\usepackage{caption} 
\usepackage{graphicx}
\usepackage{subfig}
\usepackage{enumerate}
\usepackage{enumitem}
\usepackage{tabularx,booktabs,textcomp}
\usepackage{amsthm}

\usepackage{amsmath,amssymb,amsthm}

\usepackage{enumitem}
\makeatletter 
\newcommand{\linebreakand}{%
  \end{@IEEEauthorhalign}
  \hfill\mbox{}\par
  \mbox{}\hfill\begin{@IEEEauthorhalign}
}
\makeatother 
\usepackage{amsmath}
    
\begin{document}

\title{Advancements in Enhancing Resilience of Electrical Distribution Systems: A Review on Frameworks, Metrics, and Technological Innovations}

\author{{Divyanshi Dwivedi, Sagar Babu Mitikiri, K. Victor Sam Moses Babu, Pradeep Kumar Yemula, Vedantham Lakshmi Srininvas, Pratyush Chakraborty, Mayukha Pal}

\thanks{(Corresponding author: Mayukha Pal)}

\thanks{Mrs. Divyanshi Dwivedi is a Data Science Research Intern at ABB Ability Innovation Center, Hyderabad 500084, India, and also a Research Scholar at the Department of Electrical Engineering, Indian Institute of Technology, Hyderabad 502205, IN.}
\thanks{Mr. Mitikiri Sagar Babu is a Data Science Research Intern at ABB Ability Innovation Center, Hyderabad 500084, India, and also a Research Scholar at the Department of Electrical Engineering, Indian Institute of Technology (ISM), Dhanbad 826004, IN.}
\thanks{Mr. K. Victor Sam Moses Babu is a Data Science Research Intern at ABB Ability Innovation Center, Hyderabad 500084, India and also a Research Scholar at the Department of Electrical and Electronics Engineering, BITS Pilani Hyderabad Campus, Hyderabad 500078, IN.}
\thanks{Dr. Pradeep Kumar Yemula is an Assoc. Professor with the Department of Electrical Engineering, Indian Institute of Technology, Hyderabad 502205, IN.}
\thanks{Dr. Vedantham Lakshmi Srinivas is an Asst. Professor with the Department of Electrical Engineering, Indian Institute of Technology (ISM), Dhanbad 826004, IN.}
\thanks{Dr. Pratyush Chakraborty is an Asst. Professor with the Department of Electrical and Electronics Engineering, BITS Pilani Hyderabad Campus, Hyderabad 500078, IN.}
\thanks{Dr. Mayukha Pal is with ABB Ability Innovation Center, Hyderabad-500084, IN, working as Global R\&D Leader – Cloud \& Analytics (e-mail: mayukha.pal@in.abb.com).}
}

\maketitle
\thispagestyle{empty}
\begin{abstract}

This comprehensive review paper explores power system resilience, emphasizing its evolution, comparison with reliability, and conducting a thorough analysis of the definition and characteristics of resilience. The paper presents the resilience frameworks and the application of quantitative power system resilience metrics to assess and quantify resilience. Additionally, it investigates the relevance of complex network theory in the context of power system resilience. An integral part of this review involves examining the incorporation of data-driven techniques in enhancing power system resilience. This includes the role of data-driven methods in enhancing power system resilience and predictive analytics. Further, the paper explores the recent techniques employed for resilience enhancement, which includes planning and operational techniques. Also, a detailed explanation of microgrid (MG) deployment, renewable energy integration, and peer-to-peer (P2P) energy trading in fortifying power systems against disruptions is provided. An analysis of existing research gaps and challenges is discussed for future directions toward improvements in power system resilience. Thus, a comprehensive understanding of power system resilience is provided, which helps in improving the ability of distribution systems to withstand and recover from extreme events and disruptions.

\end{abstract}


\section{Introduction}
\label{section:Intro}

    The significance of maintaining a reliable and resilient electrical infrastructure is important for meeting the expectations of consumers, utility companies, and society at large. A continuous and uninterrupted supply of power is essential for a wide range of critical loads, which includes industrial production, national security, trade, public transportation, hospital operations, and communication networks. However, the electric distribution systems are increasingly vulnerable to disruptions caused by the frequent occurrence of natural disasters and the threat of evolving cyber-physical attacks \cite{Main_percolation}. These vulnerabilities result in power outages, which are triggered by extreme weather events or malicious actions. Further, these outages result in substantial economic losses, pose threats to public safety, and disrupt essential services \cite{Dd_resilience}. This review paper provides a comprehensive exploration of the fundamental concepts of power system resilience. The primary focus is on the establishment of standardized, unified notations for power system quantitative resilience metrics. We have also addressed the need to enhance the resilience of electric distribution systems during disruptive events. The main contributions of the work are as follows:

\begin{enumerate}
    \item A clear and comprehensive understanding of the concept of power system resilience given by various organizations is provided. Different resilience frameworks available for assessing power system resilience are discussed in detail.
    \item A thorough exploration of the quantitative measures for evaluating the power system resilience metrics is discussed and compared. Also, an investigation of the relevance of complex network theory in the context of power system resilience is carried out.
    \item An essential aspect of this review is the discussion on data-driven methods used in power systems, emphasizing its role in resilience assessment and predictive analytics. 
    \item Recent techniques for enhancing power system resilience are analyzed, including both planning and operational approaches. Detailed discussions on microgrid deployment, renewable energy integration, and peer-to-peer energy trading are presented.
    \item A detailed analysis of the research gaps, challenges, and future directions is presented based on the existing literature review.
\end{enumerate}

\begin{figure*}
    \centering
    \includegraphics[scale=0.238]{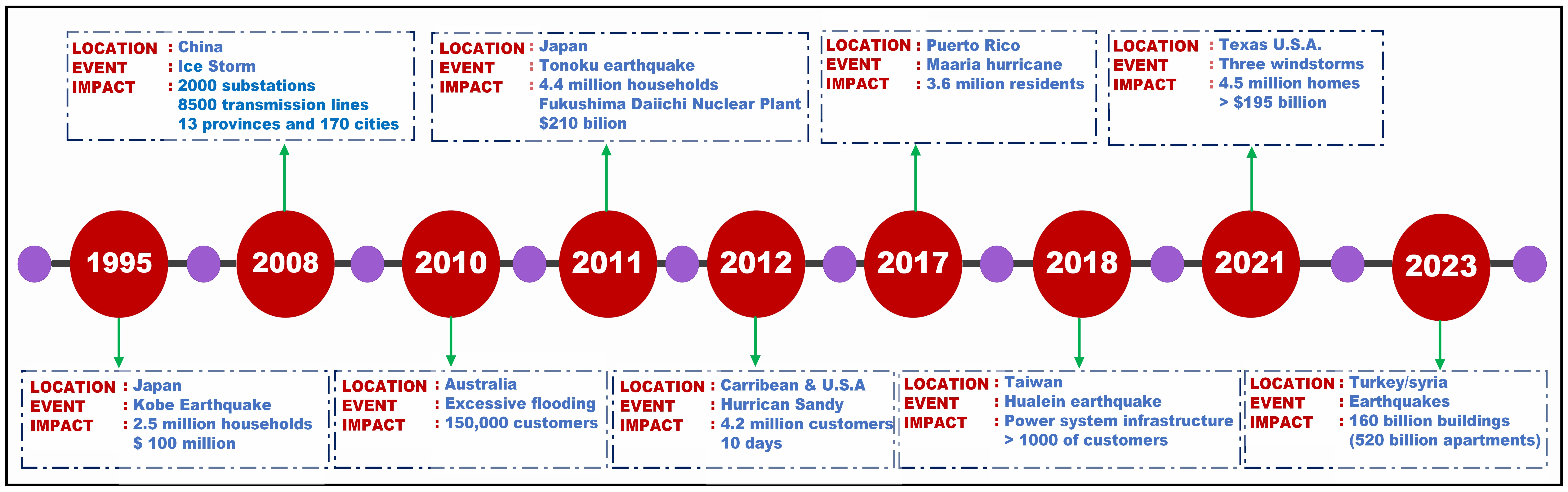}
    \caption{Chronology of most significant blackouts that caused a huge loss in infrastructure and electricity supply.}
    \label{fig:blackouts}
\end{figure*}

The paper is structured as follows: Section \ref{section:Intro}, the paper starts with a discussion on extreme events that have an impact on the resilience of systems, while also making a clear distinction between reliability and resilience. Section \ref{section:definition},  provides an extensive review of resilience metrics introduced by various organizations and a visual representation of the characteristics of resilience through triangular and trapezoidal curves. Section \ref{section:framework} elaborates on the proposed framework necessary for the development and organization of resilience metrics. Subsequently, in Section \ref{section:input}, the input requirements that are essential for assessing the system's resilience are presented. The metrics proposed in the literature are explored in Section \ref{section:metrics}, with a focus on quantitative metrics. In Section \ref{section:network}, various network-based metrics are discussed. Section \ref{section:data_driven} explores the role of data-driven methods in power systems and its application in predictive analytics for resilience assessment. Planning and operational-based resilience enhancement techniques are discussed in Section \ref{section:enhancement}. Research gaps in the evaluation of resilience and proposed potential future research directions are examined in Section \ref{section:future}. Finally, conclusions on the comprehensive exploration of resilience are provided in Section \ref{section:Conclusion}.

\subsection{Extreme Events and their Negative Impacts}

One of the major challenges for power distribution systems is the impact of natural disasters, which encompass a wide range of phenomena such as storms, earthquakes, floods, and hurricanes. These events could cause extensive damage and lead to prolonged electricity outages. In the United States, the costs associated with weather-related power disruptions vary from \$25 billion to \$70 billion annually \cite{ClimateNexus_NERCReport}. An example of one such event is Hurricane Sandy in 2012, which left 4.2 million customers without power for an extended 10-day period \cite{UMUNNAKWE2021111252}. Similarly, Hurricane Maria in 2017 had a devastating impact on 3.6 million residents in Puerto Rico, and the Hawaiian earthquake in 2018 affected thousands of customers, showing the vulnerability of power supply systems \cite{maria} and \cite{hawai}.

In 2010, Australia experienced extensive flooding that caused significant damage to six zones of substations, utility poles, transformers, and overhead cables. This event disrupted the electricity supply to approximately 150,000 consumers. Additionally, in 2008, China faced a severe ice storm that led to the malfunction of 2,000 substations and the collapse of 8,500 transmission towers, resulting in power outages affecting 13 provinces and 170 cities \cite{grid_smarter}. These examples highlight the increasing occurrence of weather-related outages, driven by global climate change, which resulted in an annual financial loss of approximately \$140 billion in the United States \cite{140}.

It is observed from recorded data between 2000-2009 that the average occurrence of weather-related events globally is around 6.7 events every year. However, this number surged to 23.2 events per year during the last five years (2019-2023). Over four decades, from 1980 to 2023, storms that accounted for the highest number of billion-dollar events total around 210 events, while tropical cyclones that caused the most significant damages amount to approximately \$1,594.4 billion \cite{ncei}. These severe storms, cyclones, and hurricanes increasingly damage power infrastructure, with distribution grids being the most susceptible segments of the power system. About 90\% of hurricane-related outages occur in this segment of the grids \cite{anurag1}. The chronological order of all the significant blackouts with severe damages to infrastructure and disruption of power supply for days/weeks are shown in Fig. \ref{fig:blackouts}.

Emerging challenges, including the impacts of climate change and the adoption of evolving technologies, further show the importance of developing resilient power distribution systems. To address these challenges, proactive actions based on empirical patterns observed in low-frequency but high-consequence events could help mitigate damages and guide utilities and policymakers in making informed decisions to enhance the resilience of the power grid \cite{toward}.

\begin{table*}[h]
\centering
\caption{Distinguishing factors between reliability and resilience in electrical distribution systems.}
\renewcommand{\arraystretch}{1.6}
\begin{tabular}{|m{0.15\linewidth}|m{0.35\linewidth}|m{0.35\linewidth}|}
\hline
\textbf{Parameters} &\textbf{Reliability} & \textbf{Resilience} \\
\hline
Predictability & Reliability often deal with regular, expected events and faults, allowing for scheduled maintenance and preparedness for common issues. & Resilience pertains to the capability to respond and recover from unexpected and extreme events, including natural disasters and cyber-attacks, requiring adaptive measures for unforeseen scenarios. \\
\hline
Performance Metrics & Reliability typically measure uptime, mean time between failures, and availability. & Resilience encompasses recovery time objectives, restoration efficiency, and adaptability under stress, focusing on minimizing downtime and restoring service quickly. \\
\hline
Dependency on Infrastructure & Reliability may rely heavily on fixed infrastructure and standard operation procedures. & Resilience often necessitates flexible infrastructure, redundancy, and diversified operational strategies, allowing for rapid response and adaptation. \\
\hline
Mitigation Strategies & Reliability emphasizes preventive measures, redundancy, and robust design to minimize failures. & Resilience prioritizes rapid response, adaptive measures, and system flexibility to recover swiftly from disruptions and unforeseen circumstances. \\
\hline
Investment & Reliability measures often require consistent, ongoing investments in maintenance and standard infrastructure. & Resilience may require intermittent but substantial investments in adaptive technology, preparedness, and response mechanisms for diverse and extreme scenarios. \\
\hline
\end{tabular}
\label{tab:compare_resilience_reliability}
\end{table*}

\subsection{Reliability v/s Resilience}

In electrical distribution systems, reliability and resilience represent two distinct yet interconnected aspects of system performance. Reliability primarily focuses on the ability of the system to deliver power without interruptions or disruptions during routine, high-probability events such as minor faults, scheduled maintenance, or voltage fluctuations. This concept emphasizes preventative measures, redundancy in components, and consistent power availability to meet everyday consumer demands. Metrics like SAIDI and SAIFI quantify the frequency and duration of outages and are often used to assess reliability \cite{HUSSAIN201956}.

On the other hand, resilience addresses low-probability, high-impact events that could severely disrupt the electrical distribution system, such as natural disasters or major equipment failures. Resilience shifts the focus from preventing such events to effectively recovering from them. It involves adaptive strategies, including backup power sources, emergency response plans, and system reconfiguration to accommodate damage or changes in circumstances. Unlike reliability, resilience is challenging to quantify, often requiring a qualitative assessment of the system's capability to cope with a range of extreme scenarios. Thus, it is important to distinguish resiliency from the reliability of power systems as detailed in Table \ref{tab:compare_resilience_reliability}.

In practice, a well-designed and planned electrical distribution system aims to have a balance between reliability and resilience, ensuring consistent service under normal conditions and the ability to recover swiftly from rare, catastrophic events. This balanced approach safeguards both everyday power needs and the system's capability to withstand and bounce back from unexpected and severe disruptions, thereby guaranteeing the continuity and stability of the electrical supply \cite{mishra}.

\section{Background of Power System Resilience}
\label{section:definition}

\subsection{Definitions of Resilience}
\subsubsection{General Definitions}
The term "resilience" is a broad and multifaceted concept used differently in various specialized fields, including engineering, organizational management, economics, psychology, biology, etc., \cite{ghosh2022comprehensive}. The word resilience is rooted in the Latin word ``resilio" which signifies the ability of an object to revert back to its original state after experiencing stress like bending, compression, or stretching \cite{grid_smarter}. Initially, in 1973, C.S. Holling introduced the concept of resilience, defining it as a measure of a system's persistence and its ability to absorb changes and disturbances while maintaining consistent relationships between populations or state variables \cite{holling1973resilience}. In 2002, Lachs et al. initiated efforts aimed at restoring and preserving the stability of the power grid in adverse conditions \cite{lachs}. Between 2002 and 2016, most research primarily focused on enhancing the power grid's reliability to ensure its stability. However, in contrast to reliability, the concept of resilience is still in its early stages of exploration. Multiple definitions exist, but a universally accepted one remains elusive. Resilience encompasses concepts such as risk assessment, reliability, recovery, and robustness, all contributing to various interpretations of the term. 

\begin{table*}
\centering
\caption{Resilience definitions proposed in UK and US by various organizations.}
\label{tab:resilience-definitions}
\centering
\renewcommand{\arraystretch}{1.5}
\begin{tabular}
  {|p{5cm}|p{8cm}|p{1.2cm}|}
    \hline
\textbf{Organization}\vspace{0.1cm}& \textbf{Resilience Definition and Attributes} \vspace{0.1cm}& \textbf{Reference}\vspace{0.1cm} \\
    \hline
\multicolumn{3}{|c|}{\textit{\textbf{Definitions by organizations in UK}} } \\
    \hline 
U.K. Cabinet Office & Resilience encompasses reliability and extends to include resistance, redundancy, response, and recovery as integral components. & \cite{uk} \\
United Kingdom Energy Research Center & The ability of a system to endure and consistently provide cost-effective services in the wake of extreme events, with an emphasis on recovery and the use of alternative methods to deliver post-disaster services. & \cite{25} \\
The UK Government Resilience Framework & Prioritizes preventing risks where possible but acknowledges the need to enhance readiness and response capabilities due to the unpredictable nature of certain crises, emphasizing a shift from solely dealing with emergencies to focusing on preparation and prevention. & \cite{uknew} \\

\hline
\multicolumn{3}{|c|}{\textit{\textbf{Definitions by organizations in US} }} \\
    \hline

U.S. Department of Homeland Security & Resiliency encompasses the capability to endure and recover from intentional attacks, accidents, or naturally occurring threats or incidents. & \cite{dhs} \\
Executive Office of the US President & Economic advantages of a more resilient power grid. & \cite{white_paper} \\
Sandia National Laboratories & Emphasizes the probability of consequences given specific threats. & \cite{sandia} \\
Los Alamos National Laboratory & Decision support tool for designing resilient systems, utilizing fragility models and criteria-based assessment. & \cite{lanl} \\
RAND Corporation & Focuses on minimizing load losses during events, emphasizing factors like redundancy and automation. & \cite{rand} \\
United States Presidential Policy Directive-21  & Infrastructure with resilience demonstrate the ability to flexibly respond to evolving conditions and swiftly regain functionality following disruption in the face of unforeseen contingencies. & \cite{15} \\
U.S. Department of Energy & Introduce a NAERM that integrate long term planning of energy supply and situational awareness in real time. & \cite{16} \\
Pacific Northwest National Laboratory  & Ability to prepare for and adapt to changing conditions, withstand disruptions, and recover rapidly from various disruptive events. & \cite{18} \\
National Renewable Energy Laboratory  & Focuses on enhancing the resilience of electrical distribution systems using model predictive control for the restoration of critical loads. & \cite{19} \\
Electric Power Research Institute & Resilience definition consists of three components: prevention, recovery, and survivability. & \cite{23} \\

National Association of Regulatory Utilities Commissioners & Emphasizes the robustness and recoverability characteristics of resilient utility infrastructure operations to reduce interruptions. & \cite{25} \\

National Infrastructure Advisory Council & The capability to absorb, adapt to, and swiftly recover from a compromised state, encompassing the ability to sustain critical functions, handle crises, and efficiently restore normal operations. & \cite{27} \\
North American Reliability Corporation & Resilience is the temporal dimension of reliability, as defined within the framework of an appropriate reliability threshold. & \cite{27} \\
North American Transmission Forum & Ability of the system and its components to minimize damage and improve recovery from non-routine disruptions in a reasonable timeframe. & \cite{32} \\
\hline
\end{tabular}
\end{table*}

\begin{figure}
    \centering
    \includegraphics[scale=0.34]{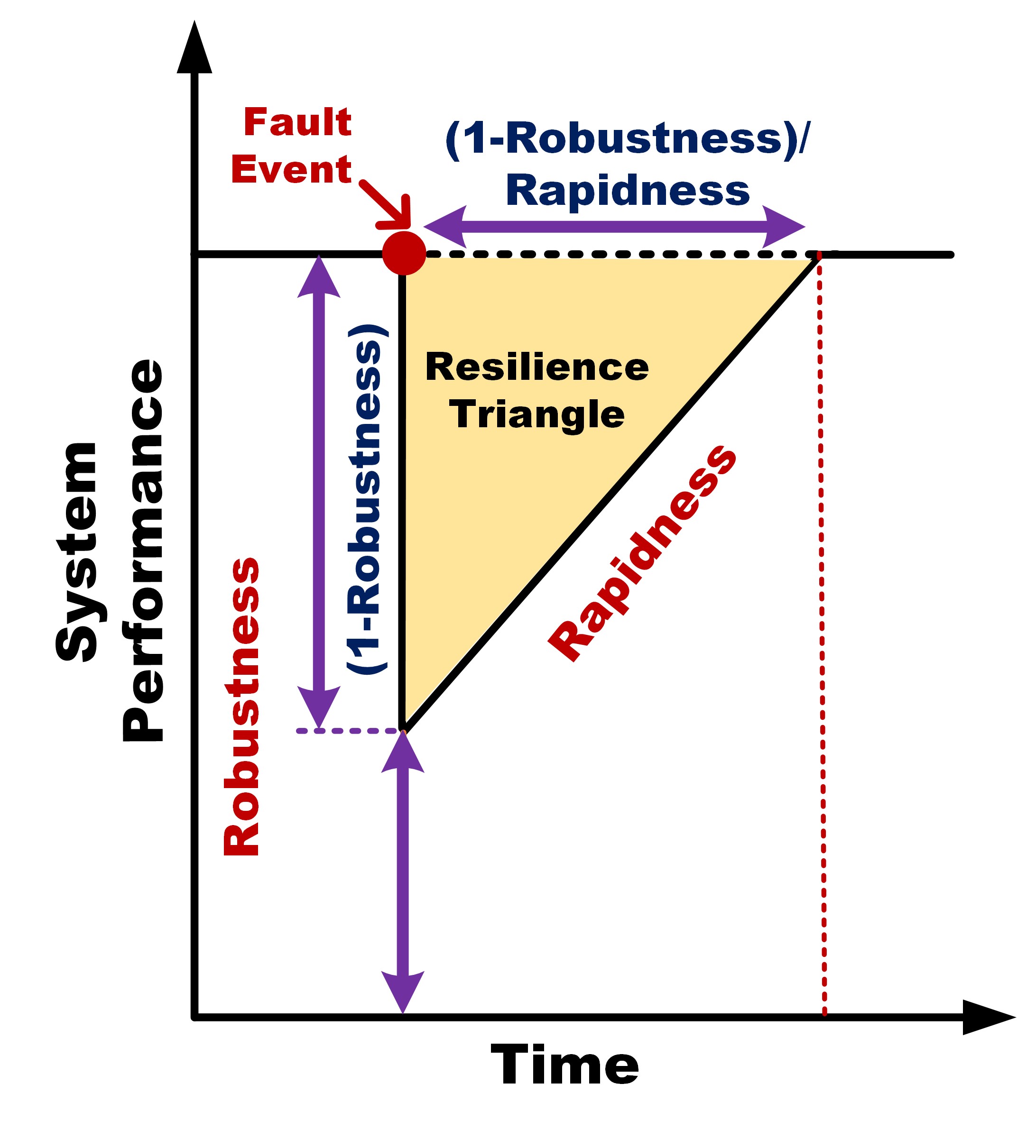}
    \caption{Resilience triangle.}
    \label{fig:triangle}
\end{figure}

\begin{figure}
    \centering
    \includegraphics[scale=0.34]{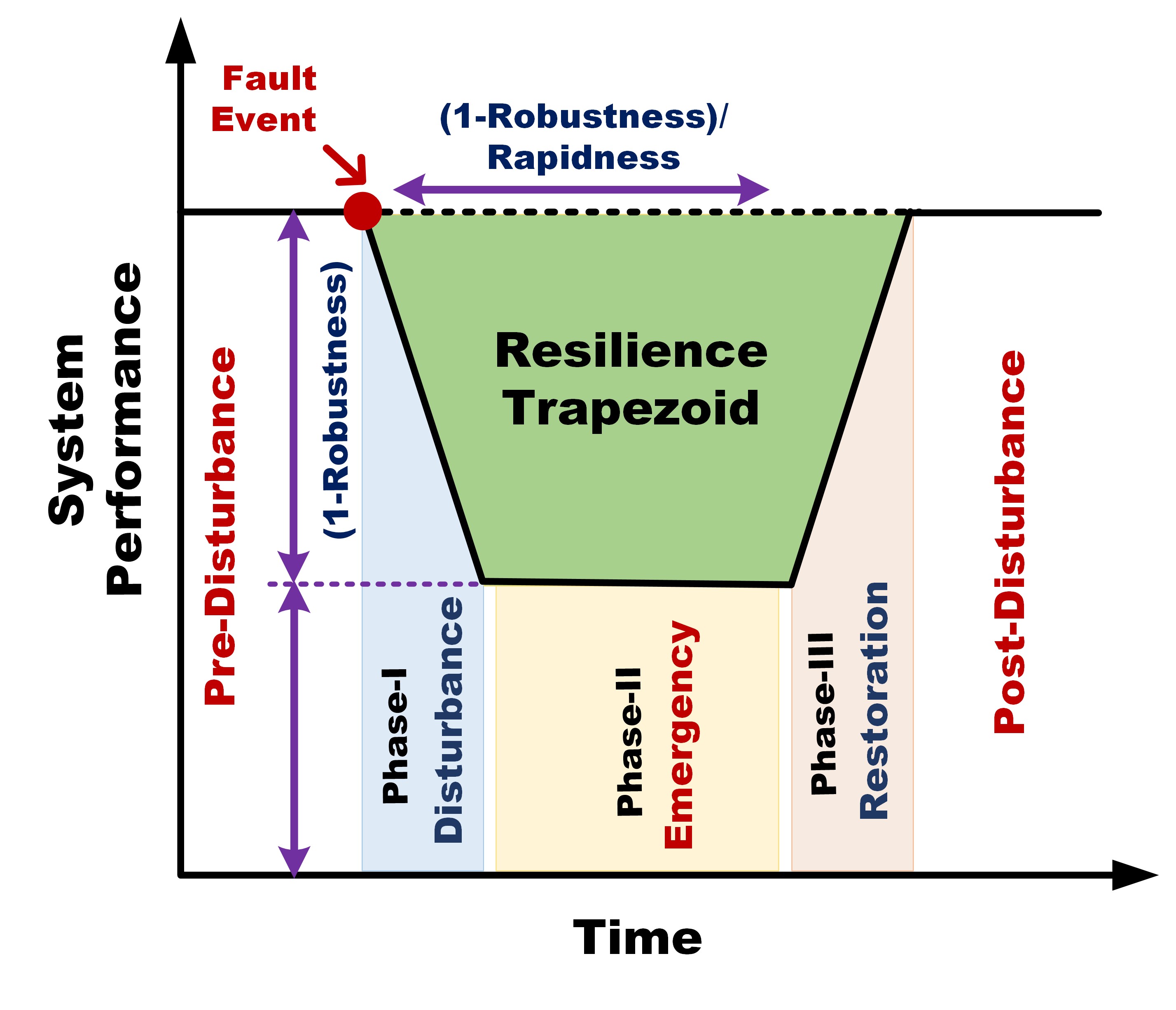}
    \caption{Resilience trapezoid.}
    \label{fig:trapezoidal}
\end{figure}

\begin{figure*}
    \centering
    \includegraphics[scale=0.24]{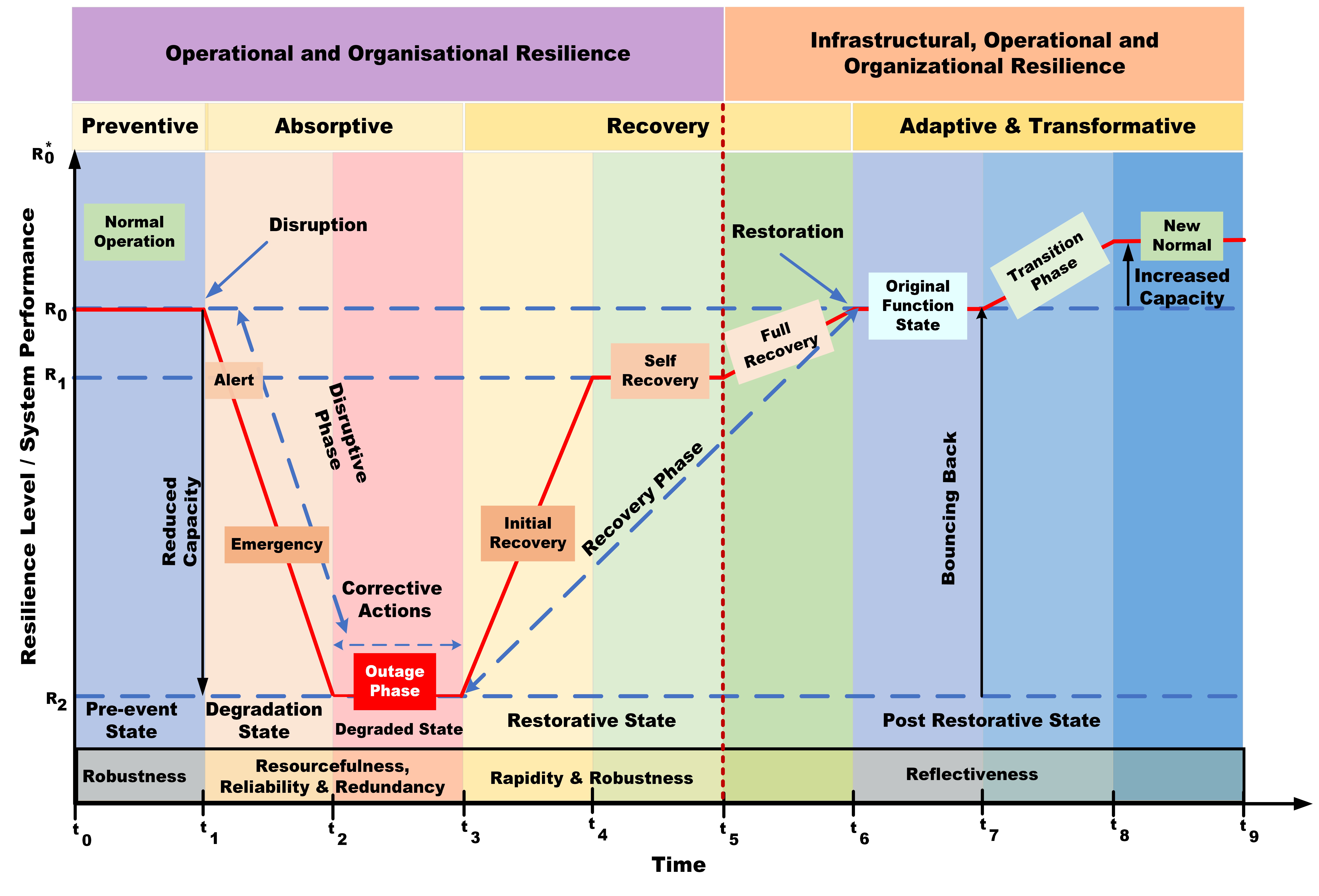}
    \caption{The resilience trapezoid depicting the power system's performance at various time points illustrating the system's transition from normal operation through disruptive events, emergency states, and eventually complete system restoration.}
    \label{fig:conceptual_trap}
\end{figure*}

\subsubsection{Definitions proposed by UN and UK}
Various entities within the power and energy engineering sectors have undertaken initiatives to define the concept of resilience and differentiate it from the notion of reliability. According to \cite{risk}, a system's resilience depends on its ability to minimize the scale and duration of disturbances' impacts. In 2009, the United Nations defined resilience generally as the capability of the system, society, or community when subjected to hazards to effectively withstand, absorb or adapt and reinstate from such hazardous impacts by maintaining and restoring back to its fundamental structures and functions \cite{kahan2009operational}. The United Kingdom Energy Research Center (UKERC) defined power system resilience as the ability to withstand disruptions, specifically to recover rapidly from failures and being able to continuously provide economical power services to customers \cite{chaudry2011building,hamidieh2022microgrids} The U.K. Cabinet Office stated that resilience encompasses reliability and extends to include resistance, redundancy, response, and recovery as integral components \cite{uk}. In 2022, ``UK Government Resilience Framework," released by the Cabinet Office, describes the strategy to fortify the systems and capabilities underpinning collective resilience \cite{uknew}. This policy outlines the proposed measures and strategies aimed at enhancing the UK's overall preparedness to manage and respond to various crises and unforeseen events. It serves as a comprehensive guide detailing how the government plans to enhance the nation's resilience in the face of diverse challenges, emphasizing the strengthening of systems to support collective resilience efforts across the board.

\subsubsection{Definitions proposed by US}
The U.S. Department of Homeland Security (DHS) \cite{dhs} defines resiliency as the ability to withstand and bounce back from accidents, attacks, or natural events. A white paper which is the government's official and authoritative report, was published by the Executive Office of the U.S. President in August 2013; it discussed the economic benefits of a more resilient power grid \cite{white_paper}. This white paper is built upon a policy framework introduced in 2011 that introduced a four-pillar conceptual strategy for the modernization of the electrical grid. The main focus of this strategy was enhancing the grid's resilience and reducing its susceptibility to weather-related disruptions. In 2015, Sandia National Laboratories introduced a resilience framework, emphasizing the probability of consequences given specific threats \cite{sandia}. Resilience, ideally, combines responsiveness to withstand and recover from disruptions with proactive measures to predict or prevent future incidents. They have also proposed a resilience analysis process (RAP) that could be used to assess baseline resilience and evaluate resilience improvements. In 2012, Los Alamos National Laboratory offered a decision support tool for designing resilient systems, utilizing fragility models and criteria-based assessment \cite{lanl}. RAND Corporation outlines a framework focusing on minimizing load losses during events, emphasizing factors like redundancy and automation \cite{rand}. In 2013, resilience was defined by the United States Presidential Policy Directive-21 (PPD-21) \cite{15} within the context of critical infrastructure. According to this definition, the resilience of critical infrastructures has the capability to adapt and withstand varying situations and promptly recover from a disruption condition during any extreme event. Additionally, the National Renewable Energy Laboratory (NREL) has incorporated the PPD-21 definition into their work, which focuses on improving distribution system resilience using model predictive control (MPC) for critical load restoration \cite{19}. It's important to note that these national laboratories are part of the Grid Modernization Laboratory Consortium (GMLC) \cite{20}. A collaborative effort between the Department of Energy (DOE) and national laboratories is aimed at advancing grid modernization. The consortium adheres to a consistent resilience definition across all studies conducted under the Grid Modernization Initiative (GMI) \cite{21}. This definition is stated as the ability to anticipate, prepare for, and adapt to changing conditions and withstand, respond to, and rapidly recover from disturbances through flexible and comprehensive planning alongside technical solutions \cite{22}.  In \cite{30}, Haimes introduces the grid's ability to autonomously return to a standard and dependable operational state with minimal human intervention. The definition from \cite{31} was employed to measure improvements in resilience. The North American Transmission Forum (NATF) defines resilience as the capability of the system and its elements to reduce damage and enhance recovery from unexpected disruptions within a reasonable timeframe \cite{32}. From the power system's perspective, \cite{33} defines resilience as the system's capability to resist high-impact, low-probability (HILP) events, swiftly recover from such events, and adjust its functioning and configuration to alleviate potential future impacts.

\subsubsection{Definition by IEEE Task Force}

After considering the similarities and differences in the definitions of resilience offered by various organizations, government, and industry bodies, the IEEE task force proposed the following definition in 2022 \cite{IEEE_Task}:
``Power system resilience is the ability to limit the extent, system impact, and duration of degradation in order to sustain critical services following an extraordinary event. Key enablers for a resilient response include the capability to anticipate, absorb,
rapidly recover from, adapt to, and learn from such an event. Extraordinary events for the power system may be caused by natural threats, accidents, equipment failures, and deliberate physical or cyber-attacks.”

The statement defines the concept of power system resilience, which refers to the power grid's capability to withstand and respond to adverse circumstances, such as disasters, accidents, equipment malfunctions, or deliberate attacks, in a way that ensures the continuation of supply to essential or critical loads which include industrial production, national security, trade, public transportation, hospital operations, and communication networks. The key elements of the power system resilience definition given by IEEE task force are:

\begin{itemize}
    \item Limiting the Extent: This means minimizing the spread or scope of damage or disruption, preventing it from affecting a larger portion of the power system.
    \item Limiting System Impact: Resilience aims to reduce the overall impact on the power system. This includes mitigating disruptions and ensuring that the grid continues to function as smoothly as possible.
    \item Limiting Duration of Degradation: Resilience efforts aim to decrease the time that critical services are affected. Shortening the period of disruption is crucial to maintaining essential functions.
\end{itemize}

\subsubsection{Inferences from Definitions}
The definitions proposed by various organizations are tabulated in Table \ref{tab:resilience-definitions}. As there is no universally agreed-upon definition for power system resilience, its relevance relies on the particular issue under consideration. However, by standardizing the key characteristics of power system resilience, such as the capability to anticipate, adapt, and recover from disruptive events, it becomes possible to standardize metrics for quantifying resilience improvement techniques and guiding resilience-oriented investments. Proposing a ``one size fits all" resilience metric remains challenging due to the diverse nature of resilience-oriented studies and predefined resilience goals \cite{anurag2}. It's important to acknowledge that events with low probability but high potential impact could result in significant damages. However, preparing for such events may be impractical and economically burdensome, especially when they are hard to predict or forecast far in advance, making it challenging for utilities and citizens to adequately prepare for potential disasters. These definitions all share a common theme, emphasizing the need for modern society's infrastructure to possess the ability to ``WRAP." This acronym underscores four key attributes \cite{anurag2}: the capability to withstand sudden adverse weather conditions or human attacks, the ability to respond swiftly to restore community balance, adaptability to new operating conditions while maintaining smooth functionality, and the importance of predicting or preferably preventing future attacks based on historical patterns or reliable forecasts.

\subsection{Characteristics of Resilience}
\subsubsection{Resilience triangle}
There is a lack of research studies that thoroughly explore the characteristics of resilience. Researchers explore various ways in which a state of resilience is achieved. However, some studies \cite{191,201,Kröger2019} have introduced the concept of the resilience triangle, using only two distinct states; robustness, and rapidness, as depicted in Fig. \ref{fig:triangle}. Robustness quantifies the system's capability by measuring the minimum functionality required and rapidness pertains to the ability to achieve prioritized objectives within a defined timeframe, thereby mitigating threats and preventing further disruption. Thus, the performance level of resilience is gauged as a function of both robustness and rapidness \cite{20}.

\subsubsection{Resilience trapezoid}
Moreover, a more systematic comprehension of the stages with respect to resiliency operation and the resilience trapezoid was explored \cite{grid_smarter,HUSSAIN201956}. The resilience trapezoid shown in Fig. \ref{fig:trapezoidal} provides a detailed analysis of the resilience state in the face of disruptive events, shown across Phases I, II, and III. In Phase I, the event impacts the network, plunging it into a disruptive phase. Phase II necessitates emergency coordination to meet load demand, and Phase III signifies the potential for restorative actions to return to normal operation. Enhancing resilience characteristics provides significant corrective planning and operational measures, serving as crucial mechanisms to expedite the restoration of normal operations within a shortened time frame. 

Therefore, a resilient system should encompass a set of fundamental capabilities, which include preventive, absorptive, recovery, adaptive, and transformative capabilities. These capabilities represent various stages in a system's response to a disruptive event. Their aim is to improve the system's resilience ($R^*_o$) and mitigate interruption in the system operations during an event. The objective is not merely to restore the system to its original operational state, $R_o$, but to ensure that it becomes less vulnerable in the face of similar triggering events. Fig. \ref{fig:conceptual_trap} depicts a standard resilience curve, where these different capabilities appear as distinct phases. However, during any given event, various other capabilities may come into play simultaneously. The preventive capability includes the anticipative aspect, which primarily strives to keep the impact and consequence of disruption within acceptable limits. The occurrence time of an event is represented by $t_o$, and the time at which the system faces interruption is represented by $t_1$; the time gap between these instances signifies the preventive capability. A longer gap is preferable as it separates the event from its associated effects, often seen in transmission networks with substantial redundancies. If there are no instantaneous fault detection mechanisms, fault detection may lead to cascading failures. When the system deviates from its normal operation condition, the absorptive state comes into play to safeguard the important infrastructure of the system while preventing enduring damage to assets or the entire system, and this phase is known as the disruption transition. System degradation is seen as a progression from a warning stage, $R_1$, which represents an initial value of operational constraint, to the emergency condition. Eventually, the system may reach an extreme condition, $R_2$, at $t_2$, where a total blackout becomes imminent. During the degraded state, corrective and emergency actions prioritize the restoration of critical loads at $t_4$, followed by a complete system recovery. Moreover, beyond these capabilities, a resilient system also has specific qualities, which are inherent attributes of the system that serve to avert system breakdown. These qualities include:
\begin{itemize}
    \item Robustness: It signifies the system's capability to absorb shocks and continue to operate or to reduce the sensitivity of outputs.
    \item Resourcefulness: It represents the capability to adeptly manage a crisis as it develops.
    \item Redundancy: It refers to spare capability that accommodates disruptions.
    \item Rapidity: It involves the ability to swiftly restore services to their pre-event state.
    \item Reliability: It indicates the system's ability to function satisfactorily across various conditions
    \item Reflectiveness: It implies ongoing evolution and adjustment of standards guided by emerging evidence.
\end{itemize}

The purpose of organizing these resilience qualities in Fig. \ref{fig:conceptual_trap} is not to establish boundaries for when they are applicable but rather to illustrate that at different phases of disruption or response, certain qualities may take precedence over others.

\section{Resilience Frameworks}
\label{section:framework}

\begin{figure*}
  \centering
  \includegraphics[width=6.4in]{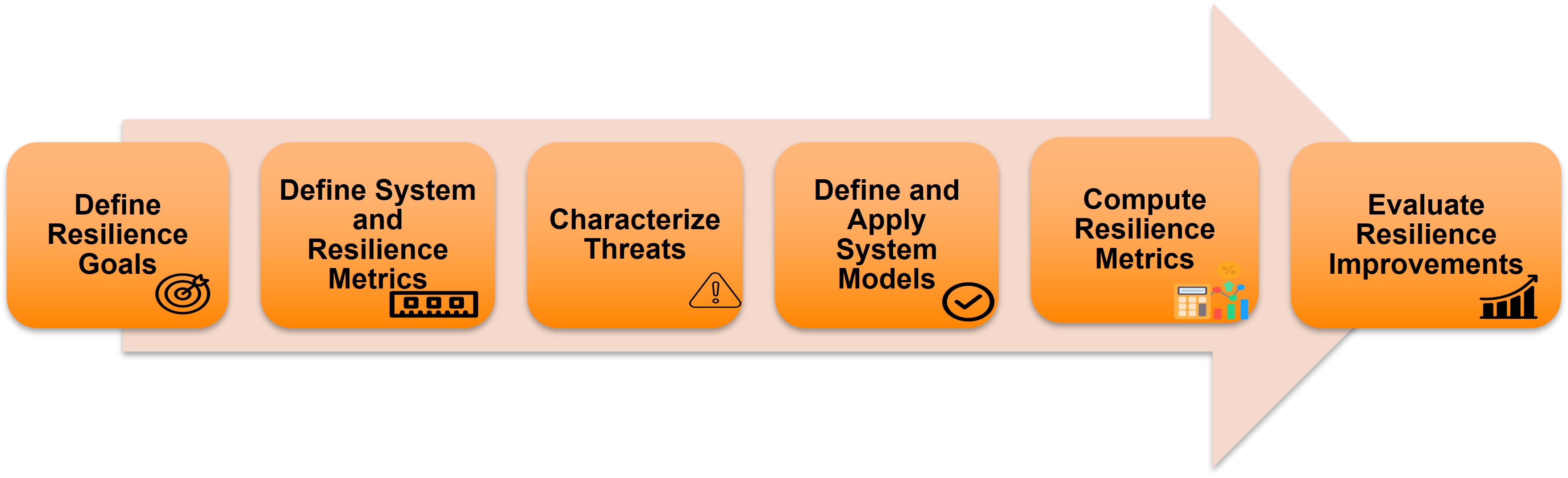}
  \caption{Illustration of the sequential stages of the resilience analysis process, providing a visual representation of the process.}
  \label{fig:process}
\end{figure*}

\subsection{Framework developed by Sandia National Laboratory}

According to the resilience analysis process outlined by Sandia National Laboratories, the development of resilience metrics should encompass several essential elements. These requirements are crucial for creating a robust framework for resilience metrics, as they must \cite{sandia}:

\begin{itemize}
    \item {Useful:} Metrics developed within this framework should be of practical utility, aiding in decision-making processes, whether conducted by humans, computational analysis, or their combination. These decisions encompass a wide spectrum, ranging from system planning and real-time operations to policy considerations.
    \item {Comparable:}  A robust metric should offer a means for effective comparison. Utilizing the same metric across diverse systems should yield valuable insights.
    \item {Applicable across Operations and Planning:} Resilience metrics must be applicable in both operational scenarios, such as preparing a system for an impending hurricane, and in planning phases, like the decision to bury electrical conductors.
    \item {Applicable and Scaleable:} These metrics should exhibit adaptability over time and geography, remaining relevant as technology advances and more intricate analytical methods emerge.
    \item {Quantitative and Qualitative:} The framework should accommodate the development of metrics that could be employed both qualitatively and quantitatively, offering versatility in assessment.
    \item {Incorporate Uncertainty:} Metrics should be populated using methods capable of quantifying the uncertainty inherent in the results.
    \item {Consider a Risk-Based Approach:} Metrics should reflect specific hazards or sets of hazards, system vulnerabilities, and potential consequences to individuals, extending beyond immediate system impacts.
    \item {Consider Recovery Time:} Resilience metrics should incorporate the duration of outages, either directly or indirectly, as a factor in their evaluation.
\end{itemize}

\begin{figure}
  \centering
  \includegraphics[width=2.8in]{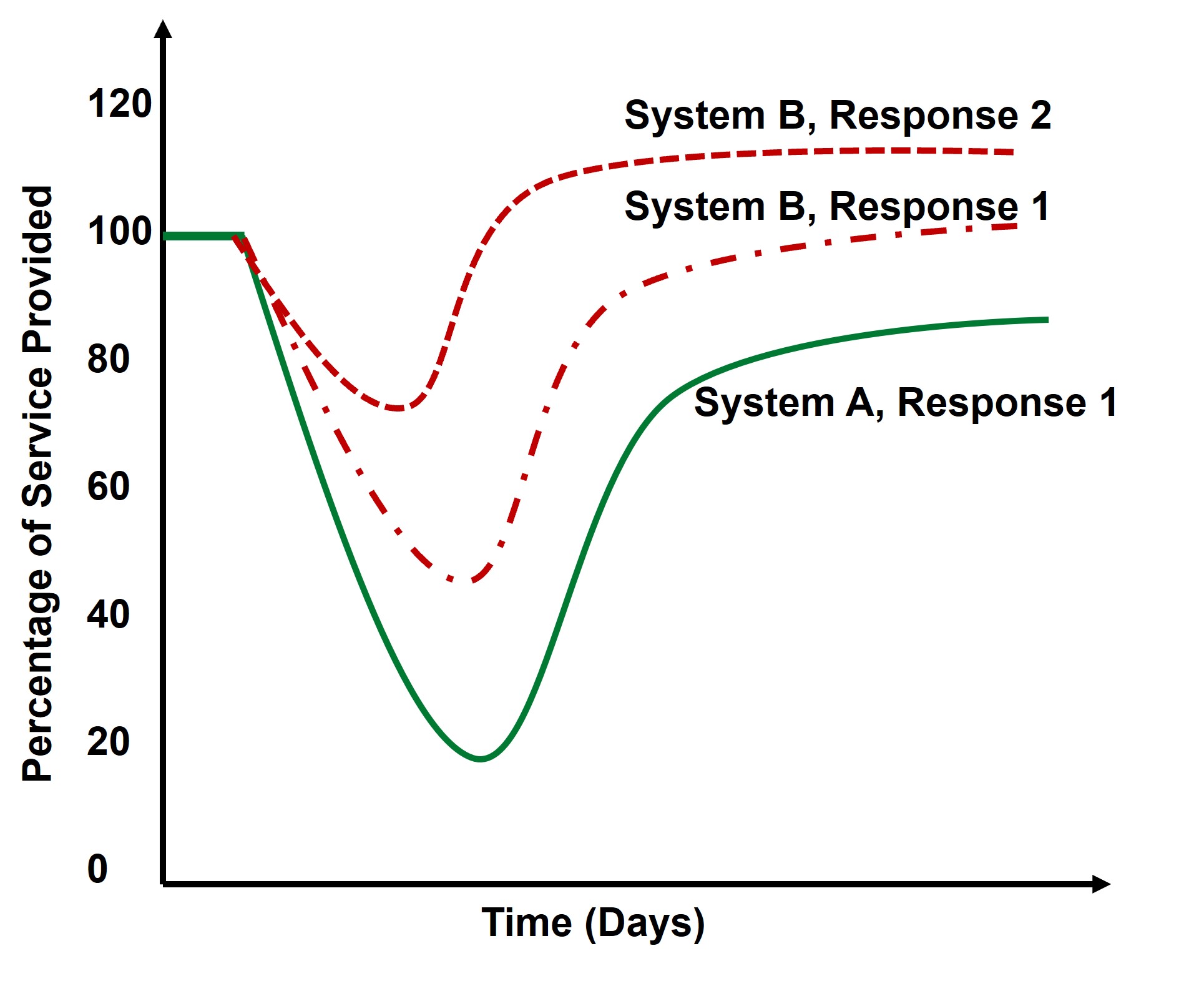}
  \caption{Resilience curve for various systems and responses proposed by RAND corporation.}
  \label{fig:rand_explnation}
\end{figure}

\begin{figure*}[t]
  \centering
  \includegraphics[width=6.4in]{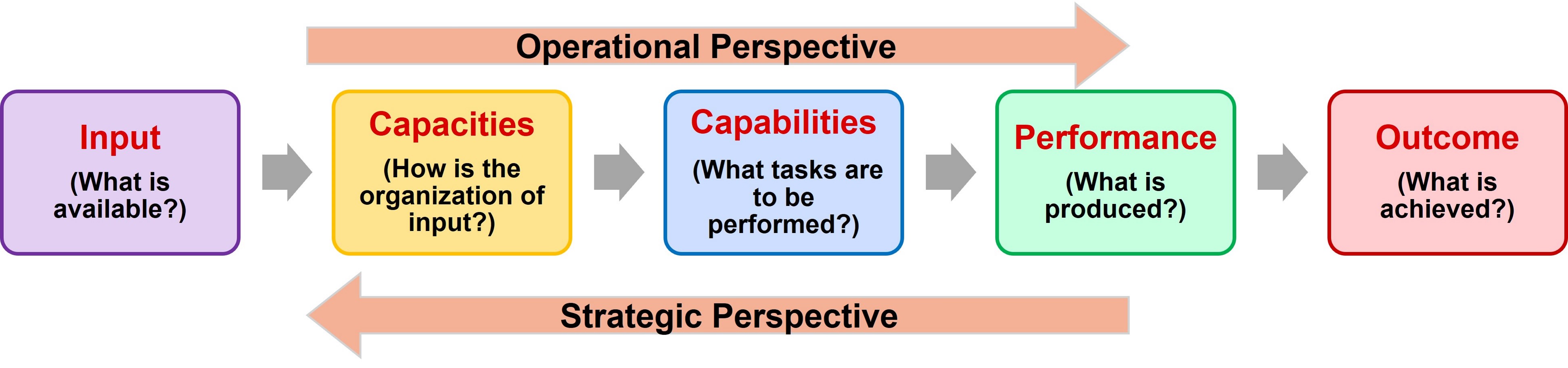}
  \caption{Resilience implementation framework by RAND corporation.}
  \label{fig:rand}
\end{figure*}

In the context of modernizing the power grid and enhancing its overall functionality, it is crucial to establish a robust framework for resilience and develop methods for quantification \cite{toward}. This resilience framework plays an important role in providing comprehensive guidelines for assessing the system's resilience. The outcomes of this assessment form the basis for making resilience-focused decisions in both operational and planning aspects of the system. Moreover, it is important to include metrics for measuring improvements in resilience, enabling a systematic evaluation of various techniques to support investment strategies.

Thus, the resilience analysis process is a structured methodology for assessing resilience and measuring the effectiveness of measures designed to improve it. The RAP consists of six primary steps designed to assess the performance of a system, as illustrated in Fig. \ref{fig:process}.
The 1\textsuperscript{st} step in the RAP process is to establish high-level resilience objectives, serving as the base for subsequent steps. Following this, the 2\textsuperscript{nd} phase involves outlining the system, establishing resilience metrics, and clearly defining the boundaries and extent of the analysis. Input from stakeholders is crucial at this stage, guiding the consideration of consequences in the analysis. The 3\textsuperscript{rd} step focuses on characterizing potential threats to the system. In the 4\textsuperscript{th} step, an estimation of the extent of system damage caused by an extreme event, identified in step three, is conducted. The 5\textsuperscript{th} step involves incorporating data regarding the impacted components into system models to assess the current state of the system. Subsequently, in the 6\textsuperscript{th} step, the outcomes derived from the system models are quantified and aligned with the resilience metrics identified in step 2. This structured approach provides a comprehensive methodology for assessing and improving the resilience of critical systems, offering valuable insights for decision-making and investment strategies.

\subsection{Framework for organizing Resilience Metrics by RAND Corporation}

A comprehensive framework is provided by the RAND Corporation, USA, for assessing resilience in energy distribution systems. The framework recognizes that resilience is not a binary state; rather, it provides multiple dimensions, including the extent of service degradation, the speed of service restoration, and the completeness of recovery \cite{rand}. The framework encompasses several key elements:

\begin{enumerate}
    \item Resilience Definition: Resilience is defined as the ability of an energy distribution system to maintain service in response to disruptions. It provides the degree of disruption across dimensions such as type, quality, time, and geography of service provision.
    \item System Design and Operation: The resilience of a system is heavily influenced by how it is designed and operated. Factors like redundancy, backup plans, and recovery strategies play a crucial role in determining resilience, as shown in Fig. \ref{fig:rand_explnation}.
    \item Cost-Resilience Trade-offs: Different responses to disruptions have varying levels of resilience and associated costs. For instance, investing in more efficient equipment during recovery may lead to higher resilience, as shown in Fig. \ref{fig:rand_explnation}. This highlights the importance of informed decision-making regarding resource allocation.
    \item Timescale Considerations: Resilience is not static; it changes over time. Systems that are continually maintained and upgraded may improve their resilience, while neglecting maintenance may lead to declining resilience. Long-term sustainability is a key aspect to consider.
\end{enumerate}
The framework acknowledges that terms such as reliability, robustness, recoverability, sustainability, hardness, vulnerability, fault tolerance, and redundancy are often related to resilience. However, for the specific context of assessing resilience in energy distribution systems, the focus is on capturing the relevant aspects of service delivery, system design, system operations, disruptions, costs, and timescale \cite{rand}. This framework helps in assessing progress, ensuring system resilience, and identifying enhancement techniques in terms of resilience metrics. The metrics of resilience serve different needs at different levels of decision-making. It explains how inputs contribute to desired outcomes as shown in Fig. \ref{fig:rand}. This hierarchical approach to metrics enhances our understanding of achieving outcomes effectively and efficiently \cite{rand}.

\section{Essential Inputs for Assessing Electrical Distribution System Resilience}
\label{section:input}

The resilience evaluation process comprises several key phases as shown in Fig. \ref{fig:phases}. It begins with event assessment, where the impact of a disruptive event is analyzed. Following this, component impact and fragility modeling is employed to assess how individual system components may be affected. For these two steps, accurate and diverse inputs are important to the process of quantifying the resilience of electrical distribution systems; this is discussed in detail in this section.

The subsequent phases are system response and resilience assessment, which examine the system's reaction to such events in terms of quantitative metrics and evaluate the system's overall resilience in the face of these events discussed in Sections \ref{section:metrics}, \ref{section:network}, and \ref{section:data_driven}. Finally, in Section \ref{section:enhancement}, the discussion on how resilience enhancement seeks to identify strategies and measures to strengthen the system's resilience and make it better prepared to withstand and recover from disruptive events.

\subsection{Fragility Models}

\begin{figure}[b]
    \centering
    \includegraphics[scale=0.5]{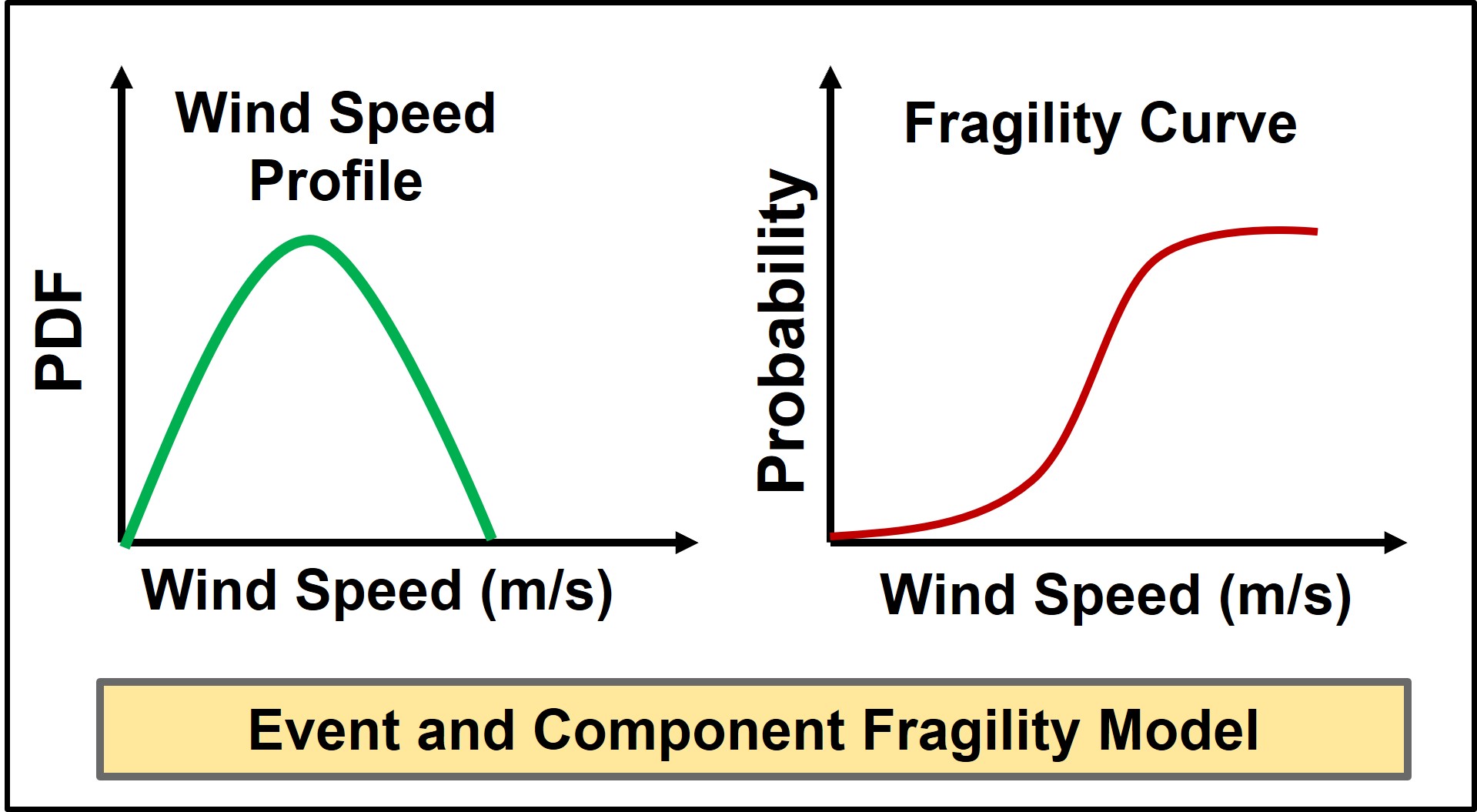}
    \caption{Visual representation of a fragility curve for an extreme wind event. The plot on the left displays a probability distribution function illustrating wind speed variations during wind events. On the right, the fragility curve depicts the likelihood of disruptions relative to wind speed.}
    \label{fig:fragility}
\end{figure}

In assessing the resilience of electrical distribution systems, the component analysis assumes a main role. Certain components within the system hold more significant sway over its overall functionality, necessitating the development of metrics to show their criticality. To gauge the effects of extreme events on these components, an essential factor is the correlation between the force of an extreme event and the resultant component failure or damage as shown in Fig. \ref{fig:fragility}. This connection is embodied through the application of fragility curves, characterized by fragility function. Fragility functions show the likelihood of component failure with the intensity of a disruptive event \cite{mujjuni2023evaluation}. In the case of power system components, these fragility curves often follow a lognormal distribution and should be customed to the particular site. Constructing fragility curves shows the collection of data like wind speed or damage levels as shown in Fig. \ref{fig:fragility} and typically employs two primary methodologies:

\begin{figure}
  \centering
  \includegraphics[width=3.6in]{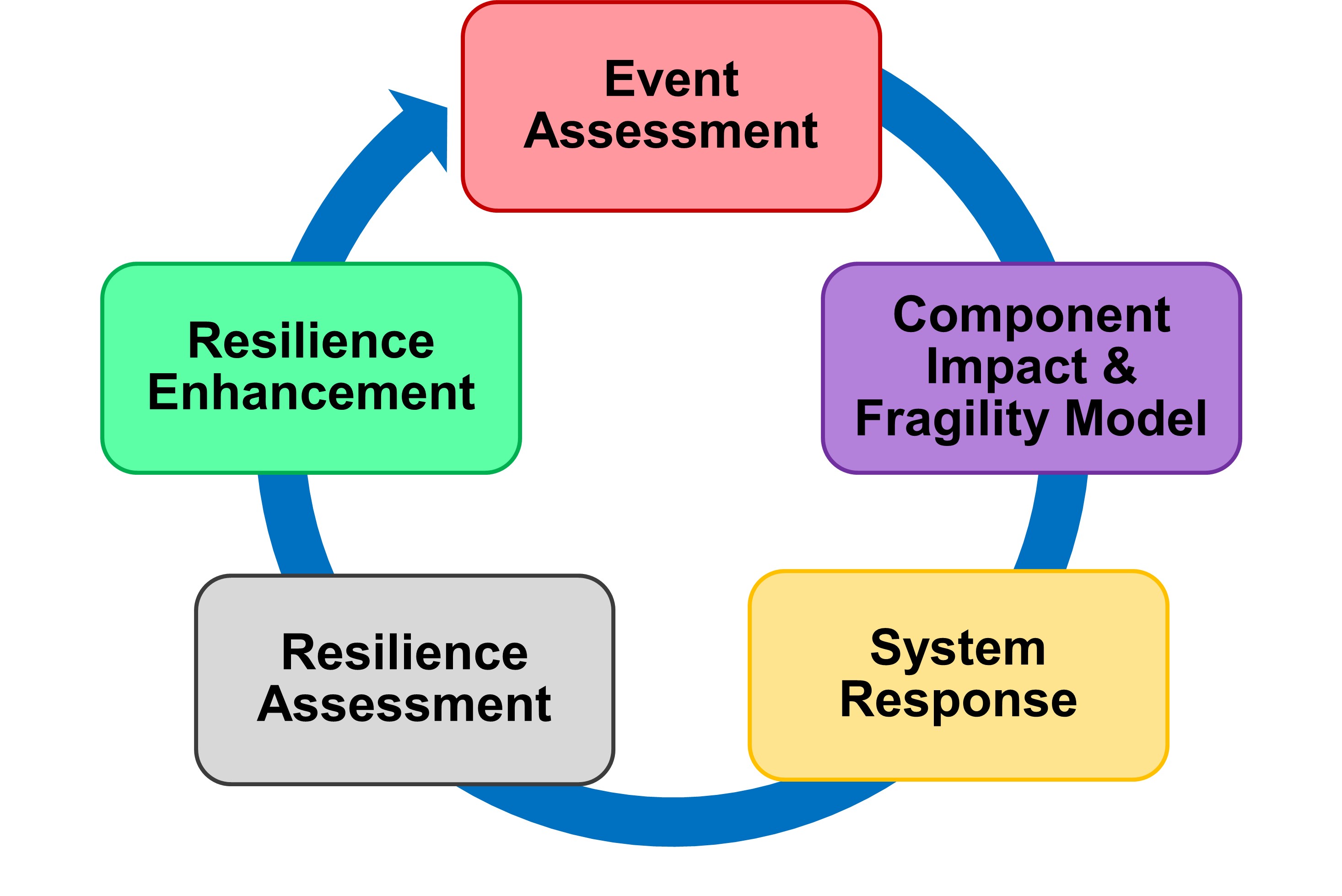}
  \caption{Resilience assessment framework.}
  \label{fig:phases}
\end{figure}

\subsubsection{Statistical Models}

Statistical models rely on techniques that include additive models, generalized linear, and accelerated time-based failure models. These models use historical data to predict component failures based on the characteristics of disruptive events, employing statistical methodologies \cite{frag1}.

\subsubsection{Simulation Models}

In contrast, simulation models employ computer or physical simulations to replicate authentic event scenarios. For instance, to simulate the effect of wind speed on a transmission line, these models expose the line to varying wind speeds, thus capturing realistic damage levels \cite{frag1}. For a deeper exploration of fragility functions and fragility curves specific to power systems, further references offer extensive insights \cite{frag3}.

\subsection{Data for Resilience Assessment}

Data requisite for quantifying the resilience of electrical distribution systems is taken from a range of sources, encompassing power utilities, system operators, meteorological agencies, and governmental entities. However, obtaining such information could be strenuous for a given system as the frequent occurrence of disruptive events at a given location is low. Also, the information about equipment failures or vulnerabilities in the system may be sensitive and not readily disclosed by utilities or other stakeholders.

\subsubsection{Data - Failures and Outages}

Actual data on failures and outages, comprising of variables such as failure duration and the number of impacted customers, are taken from power utilities that have faced disruptions, such as hurricanes. This data is crucial for understanding the system's response to disruptive events \cite{frag1}.

\subsubsection{Data - Disruptive Events}

Meteorological data, including information on windstorm speeds, is secured from entities like the National Oceanic and Atmospheric Administration (NOAA) and the National Hurricane Center \cite{noaa}. These organizations also furnish tools for disaster analysis, proffering real-time coastal information and historical weather maps. Additionally, utilities have made considerable investments in wildfire monitoring systems reliant on data from remote automated weather stations.
Several other resources, including the United Nations Office for Disaster Reduction, the Federal Emergency Management Agency, the Transmission Availability Data System, the Residential Energy Services Network, and the Multidisciplinary Center for Earthquake Engineering Research provide valuable data for resilience assessment \cite{mujjuni2023evaluation}.

Thus, the amalgamation of these data sources into models for power system failure facilitates a comprehensive evaluation of the system's response to disruptive events. The selection of data acquisition methods, whether analytical, experimental, empirical, judgmental, or hybrid, depends on particular circumstances, data availability, and the system's scale. Analytical techniques are often preferred, for small-scale system configurations, due to their simplicity and lower computational demands.

\section{Quantitative power system resilience metrics}
\label{section:metrics}

The field of resilience lacks universally accepted metrics and established evaluation methodologies. Even though several resilience metrics have been proposed, there remain ongoing discussions regarding the creation of a standardized framework for these metrics. In this section, several resilience metrics are discussed and categorized depending on the various features and factors. The terms metrics, indicators, evaluation parameters, and evaluation indices interchangeably for the resilience evaluation. The resilience metrics are categorized into several types depending on factors like the nature of the parameter being evaluated, the magnitude scale of the parameter evaluated, and the type of the events considered in the context of electrical and non-electrical parameters, etc.



Several research studies indicate that it is necessary to identify the various constituent capabilities to quantify the resilience \cite{nan2017quantitative, jufri2019state}. Resilience, as a dynamic concept, is not fully captured by a single indicator, as it evolves over time and is not confined to a single instance. In \cite{panteli2017metrics}, a framework is proposed consisting of four metrics compromising of all phases of the resilience trapezoid namely $\Phi\Lambda E\Pi$ (pronounced as FLEP), where $\Phi$ and $\Lambda$ specify how quickly and how much low the resilience curve drops in the disturbing state, respectively. $E$ denotes the extent of the post-degradation state and the $\Pi$ signifies how quickly the network recovers to its pre-event disturbance state. 


Assessing and comparing resilience following extreme events remains challenging even with the resilience trapezoid framework due to the resulting consequences \cite{yao2022quantitative}. Hence, it is not straightforward to find the curve corresponding to higher resilience due to different trapezoidal curves for different topologies and infrastructures. Several works have been proposed with multiple metrics for the evaluation of resilience. The commonly used metrics for evaluating resilience are elaborated through a resilience trapezoidal curve depicted in Fig.\ref{fig:conceptual_trap}, are as follows:

\subsubsection{Energy or Demand Served}  
It is the total amount of energy or the demand served from the instant the event occurred ($t_{1}$) to the completion of the restoration period where the post-event resilience curve meets the pre-event original function state ($t_{6}$). The computational values for these metrics have been found to be calculated as \cite{bhusal2020power}:

\begin{equation}
     \text{Energy served} = \int_{t_{1}}^{t_{6}}\!R(t) \;dt \qquad
\end{equation}

\begin{equation}
     \text{Demand served} = 100\left(1-\dfrac{L_{mn}}{L_{max}}\right)
\end{equation}

where $R(t)$ denotes the function of system performance, $L_{mn}$ is the observed maximum loss in the performance of the system, and $L_{max}$ is the expected load loss in case of a complete blackout. 
    
\subsubsection{Energy Not Supplied (ENS) and Expected Energy Not Supplied (EENS)}
The ENS is a variable loss index that gives the total amount of actual cumulative energy not supplied during the assessment period. ENS$_{t}$ is the instantaneous energy not supplied. The EENS is a predictive estimate or statistical likelihood of power estimate based on historical data and system modeling factors. It informs the design and planning of a dependable power system and evaluates how well it performs under various hypothetical scenarios. These energy-based parameters \cite{panteli2017metrics} are computed as follows:
\begin{eqnarray}
    \begin{split}
        \text{ENS} & = \int_{t_{0}}^{t_{6}}\!(R_{0}-R(t))t\; dt \; \big| P(R(t)>RC)\\
        & = \int_{t_{1}}^{t_{6}}\![100-R(t)] \;dt
    \end{split}
\end{eqnarray}

\begin{equation}
  \text{EENS}  = 
    \sum\limits_{t_{1}}^{t_{6}}\text{ENS}_{t} \;.\; P_{ENS_{t}} \qquad \qquad \qquad \qquad \quad
\end{equation}
where $RC$ is the reserve capacity. $P(R(t) > RC)$ indicates the probability value when the system performance is greater than $RC$. $P_{ENS_t}$ is the ENS probability value obtained from the probability distribution curve of ENS \cite{mujjuni2023evaluation}.

\subsubsection{Loss of Load Metrics} During an outage caused by severe weather conditions, these metrics quantify the amount of load lost and economic losses incurred due to the extreme events. The various metrics with their approaches for calculating these types of metrics are as follows:
\begin{itemize}
    \item \textit{Loss of Load Probability} (LOLP) - It is a metric that quantifies how often the demand for electricity surpasses the available power generation capacity \cite{sandia}.
    \begin{equation}
        \text{LOLP} = \dfrac{\sum\limits_{t} \;k_{t}\; \big| P(R(t) > A)}{t_{6} - t_{1}}
    \end{equation}
    where $A$ is the generation capacity and $k$ indicates the number of times the given condition (probability) is satisfied. $t_{6}$ is the instant when the resilience curve reaches its original function state or the instant at which the system restoration to its original state is achieved. $t_{1}$ is the disruption starting time.
    \item  \textit{Loss of Load Expectation} (LOLE) - The duration during which a load outage occurs is referred to as loss of load expectation \cite{sandia}.
    \begin{equation}
      \text{LOLE} = \frac{1}{N_{k}}\sum_{t}\;k_{t}|P(R(t)>A)  
    \end{equation}
    where $N_{k}$ is the number of considered simulations.
    \item \textit{Loss of Load Frequency} (LOLF) - It is the cumulative frequency of system \cite{mahzarnia2020review}.
    \begin{equation}
       \text{LOLF} = F(OC>RC) 
    \end{equation}
    where $OC$ is the outage capacity.
    \item \textit{Value of Loss Load} (VOLL) - It represents the monetary value associated with the power that couldn't be supplied during such outages. The overall VOLL is calculated by summing up the product of the energy not delivered at each point in time and the corresponding VOLL at that moment \cite{moreno2020reliability}. This summation extends from the initial occurrence $t_{1}$ of the outage event to the point when the power system fully recovers to its original resilience level $t_{6}$.
    \begin{equation}
        \text{VOLL} = \sum_{t=t_{1}}^{t_{6}}\text{ENS}_{t} \; . \; \text{VOLL}_{t}
    \end{equation}
    \item \textit{Outage Price} (OP) - The cost incurred by a system as a result of the disruption of a portion of its loads is termed as outage price \cite{moreno2020reliability}.
    \begin{equation}
        \text{OP} = \text{EENS}\times\text{VOLL}
    \end{equation}
\end{itemize}

\subsubsection{Value at Risk (VaR) Metrics}
Value at Risk (VaR) and conditional value at risk (CVaR) \cite{tavakoli2018cvar}, are both parameters characterized by their nature, and excel in quantifying resilience, as they effectively encapsulate the core principles of resilience analysis, which focus on extreme conditions. These metrics serve as valuable measures, focusing on the likelihood of risks surpassing specific thresholds in terms of impacts or consequences. The following example best illustrates the difference between them, a 5\% VaR of 1,000 GWh implies a 5\% probability that the energy demand to be served exceeds 1,000 GWh, while a 5\% CVaR of 1000 GWh signifies the value in average of the most significant 5\% potential energy deficits is 1,000 GWh. The formulas for the computation of these metrics are:
\begin{equation}
    \text{VaR}_{\alpha} (X)  = \text{min} \biggl\{z \in \mathbb{R} : p \bigl\{(X \leq z ) \geq \alpha \bigl\} \biggl\}; \; \alpha \in [0,1]
\end{equation}
\begin{equation}
     \text{CVaR} = E_{X}\big[X \big| X\geq \text{VaR}_{\alpha}(X)\big]
\end{equation}

where, $p$ indicates the probability of occurrence of associated scenarios, $\alpha$ indicates the level of confidence, $z$ is the threshold value, and $E_{X}[X]$ is the expectation of a random variable $X$ which is considered here as the system risk index.

\subsubsection{Degradation Metrics}
Resilience degradation metrics \cite{amirioun2019metrics} refers to measures or indicators used to assess the decline or deterioration in the resilience of a system, process, or organization. The metrics that are used to quantify and monitor the degradation level and time taken by the system to degrade in context to various challenges and disruptions. Some of these degradation indices are explained as follows,
\begin{itemize}
    \item \textit{Absorptivity} - It is also known as the vulnerability index. It is defined as the ratio of the difference between the pre ($R_{0}$) and post-event ($R_{2}$) resilience levels to the pre-event resilience level. This metric is quantified as:
    \begin{equation}
        \text{Absorptivity} = \frac{R_{0}-R_{2}}{R_{0}}
    \end{equation}
    \item \textit{Normalized Degradation Index} (NDI) - It is a metric that standardizes the resilience degradation measurement so that, it may be assessed with various systems in comparison. This index gives a normalized assessment of how resilience declines over time or in response to particular challenges, accounting for a variety of elements, scales, or environments.
    \begin{equation}
        \text{NDI} = \dfrac{\int\limits_{t_{1}}^{t_{2}}(R_{0} - R(t)) \; dt}{R_{0} (t_{2} - t_{1})}
    \end{equation}
\end{itemize}

\subsubsection{Restoration Metrics} 
These are the metrics that are helpful to evaluate the system performance in the post-event situation and during the system recovery stage ie., from $t_{3}$ to $t_{9}$. This period is also termed as recovery stage ($t_{r}$) in literature. In general, these metrics specify the recovery time, amount of the load or generation recovered, adaptability, and restoration efficiency index, etc., These restoration metrics are defined as follows;
\begin{itemize}
    \item \textit{Restoration Efficiency Index} (REI) - It is a normalized metric for determining how effectively the restoration phase operates. It represents the restoration efficiency, whose values depend not only on the restoration process but also on the distribution branch outage probabilities, the main outage scenario, and the restoration beginning point \cite{amirioun2019metrics}.
    \begin{equation}
        \text{REI} = \dfrac{\int\limits_{t_{3}}^{t_{4}}(R(t) - R_{2}) \;dt}{(R_{0} - R_{2}) (t_{2} - t_{3})}
    \end{equation}
    \item \textit{Critical Load Restoration} (CLR) - It is defined as the area under the restoration curve from the start of initial recovery time (t$_{3}$) to the start of final recovery time (t$_{4}$) \cite{gao2016resilience}.
    \begin{equation}
        \text{CLR} = \int_{t_{3}}^{t_{4}} R(t) \;dt
    \end{equation}
    \item \textit{Integrated Resilience Index} (IRI) - It is a normalized metric for computing the normalized standard for the efficiency index of restoration. Its computing consists of integrating the degradation, degraded, and recovery phases of Fig. \ref{fig:conceptual_trap}. 
    \begin{equation}
        \text{IRI} = \dfrac{\int\limits_{t_{1}}^{t_{4}}\! R(t) \; dt}{R_{0} (t_{4}-t_{1})}
    \end{equation}
    \item \textit{Restoration Economic Assessment} (REA) - This metric \cite{amirioun2019metrics} is used to assess the cost and benefit analysis, time frame discounting, uncertainty analysis and new marketing analysis, etc., The computation of this parameter involves:
    \begin{equation}
\text{REA}=\dfrac{\sum\limits_{j=1}^{N_{b}} \sum\limits_{l=1}^{n_{j}} \int\limits_{t_{3}}^{t_{6}}c_{jl} \; \cdot \; (S_{jl}-S_{jl}(t)) \; dt-C_{rep}}{\sum\limits_{j=1}^{N_{b}} \sum\limits_{l=1}^{n_{j}} \int\limits_{t_{3}}^{t_{6}}c_{jl} \; \cdot \; P_{jl} \; dt}
    \end{equation}
    where, $S_{jl}$ is the load loss at $t_{3}$, $S_{jl}(t)$ is the real load loss \textit{l} on bus $j$ at time $t$ during power restoration. $C_{rep}$ signifies the economic cost of repair works. $c_{jl}$ is the economic loss of load $l$ at bus $j$ in unit time.
    \item \textit{Total Restoration} (TR) - It encompasses all the resources expended during the recovery phase following a disaster \cite{mahzarnia2020review}. The computation of this metric is as:
    \begin{multline}
     \text{TR} = \sum_{t}\Bigg[\sum_{j} \text{LC}_{j} \times \text{WH}_{j,t} + \sum_{n} \text{CR}_{n} \times \text{PR}_{n,t} + \\
     \sum_{z} \left(\text{OP}_{t}-\text{OP}_{n} \right) \times (\text{PG}_{z,t} \text{D}_{t}) \Bigg]
\end{multline}

    where, LC$_{j}$ is the labor ($j$) hourly wages, WH$_{j,t}$ is the number of working hours for each person in category $j$ and $t^{th}$ interval. OP$_{n}$ and OP$_{t}$ are generator operating cost during regular conditions and $t^{th}$ stage. PG$_{z,t}$ is the power generated by generator $z$ during the $t^{th}$ stage. CR$_{n}$ is the cost of replacement for the pieces of $n$ categories. PR$_{n,t}$ is the number of pieces replacement is done in the $t^{th}$ repair stage of $n$ category. D$_{t}$ is the t$_{th}$ stage recovery duration.
\end{itemize}

\subsubsection{AWR metric framework}
This AWR metric framework stands for Anticipate, Withstand, and Recover. The author of the literature
\cite{kandaperumal2021awr} developed this AWR-based framework for evaluating resilience at different levels for real-world systems.
\begin{itemize}
    \item \textit{Anticipate Metric Computation} - The RWS assesses common threats such as tsunamis, volcanoes, cyclones, and earthquakes by employing the ``anticipate" metric, which takes into account the system's exposure to these threats. In this evaluation process, domain scores are assigned for each threat, and a weighted sum score is determined through a specific equation.
    \begin{equation}
      R_{A} = \sum_{i=1}^{N} T_{i} W_{i}^{T} +   \sum_{i=1}^{N} P_{i} W_{i}^{P} + \sum_{i=1}^{N} R_{i} W_{i}^{R}
    \end{equation}
    where $R_{A}$ indicates the anticipated metric score delivered by the combining scores $T_{i}$, $P_{i}$ anD $R_{i}$  with their corresponding weights $W_{i}^{T}$, $W_{i}^{P}$ and $W_{i}^{R}$ associated with the threat and vulnerability domain, power delivery and loads domain, and recovery and restoration domain respectively.
    \item \textit{Recovery and Withstand metrics for RWS} - The evaluation of these metrics is evaluated through the Analytic Hierarchy Process (AHP) for each scenario, considering both with and without the implementation of resilience investment for different levels of degradation. The AHP system weights are determined by constructing a pairwise comparison matrix (PCM), which reflects the influence of the factors on the system's resilience. Using these weights along with the linearized performance scores obtained, the resilience score is calculated as follows:
    \begin{equation}
        R_{i} = \sum_{j=1}^{N} \rho_{i,j} W_{i}
    \end{equation}
    where $\rho$ specifies the linear transformation, $i$ and $j$ indicate the alternative and criteria.
\end{itemize}
The anticipate metric mentioned above holds on the assumption that early warning threat detection is performed and the warning triggers are also available for computing the anticipation metric score.

In addition to the above-mentioned metrics, a few other metrics such as distortion rate ($\Theta$) and linepack effect ($\Delta t^{*}$) are proposed for the Integrated Power and Natural Gas Systems (IPGS) \cite{xie2022resilience}. These metrics are utilized to quantify the coupling tightness and the stability protections from natural gas transients respectively. Distortion rate models the whole process of IPGS under extreme weather conditions and the gas linepack in adjacent pipes is availed for generation of the enough electricity for gas-fired units. \cite{rocchetta2022enhancing} uses the metrics for quantification of the vulnerability. Resilience is a concept that relates to power outages and disruptions not only due to natural hazards and extreme events but also because of cyber attacks. Some of the metrics that are used to quantify the resilience affected by cyber-attacks are mean time to detect, mean time to respond, and incident response effectiveness \cite{fysarakis2023phoeni2x}. 

\section{Complex Network Theory Applications in Power System Resilience }
\label{section:network}

Various studies have utilized complex network theory to investigate the resilience of critical infrastructure. The application of complex network theory to assess the resilience of intricate physical infrastructure networks was initially proposed by R\'eka Albert and Albert-L\'aszl\'o Barab\'asi \cite{RevModPhys.74.47}. In the domain of power systems, various problem related to reliability and resilience has been solved using complex network theory which includes, critical node identification \cite{vul, euro}, restoration strategies for distribution systems \cite{restoration}, connectivity loss \cite{vul2}, damages and improvements \cite{damage}, reliability and disturbances \cite{reliability}, unserved energy/load \cite{unserved}, cascade effects \cite{cascade}, and blackout size \cite{complex}. Analyzing the resilience of distribution systems through complex network analysis should not be seen as an effort to enhance already effective reconfiguration strategies. Instead, it provides a broader perspective that facilitates a better grasp of distribution systems, improving their operational efficiency and planning. Power distribution systems are marked by the existence of densely interconnected nodes, where multiple loads depend on lengthy secondary and tertiary feeder branches. Disruptions caused by events like a terrorist attack or weather-related damage targeting these critical nodes significantly impact the continuity of power supply to consumers \cite{cired2}. 

In the initial step of examining an electric distribution system using the framework of complex network theory, a system model is used to create a graph \cite{graph1}. This graph, denoted as $G = (N,E,W)$, represents a power distribution system. It consists of a collection of nodes $N$, a set of edges $E$, where each edge is directed from a `From' node $i$ to a `To' node $j$, and a set of weights $W$ that signify the power flow through these edges. Using the obtained network various topological metrics could be computed which help in understanding the structural properties of the graph, including:

\begin{enumerate}
  \item \textit{Graph Diameter ($D_g$):} $D_g$ is defined as the maximum geodesic distance, which quantifies the number of edges in the shortest path connecting the two farthest nodes \cite{ref40}. It is expressed as:
  
  \begin{equation}
  D_g = \max \left\{ d(n_i, n_j) : \forall (n_i, n_j) \in N \right\} 
  \end{equation}
  
  where, $(n_i, n_j)$ represents pairs of nodes within the graph $G$.

  \item \textit{Aggregate Betweenness Centrality} (ACB): ACB is defined as the average difference in betweenness centrality between the most central node $n$ in the graph, which holds the highest betweenness value, and all remaining nodes. For node $n_i$, ACB is calculated as follows:
  
  \begin{equation}
  \text{ACB}(n_i) = \frac{1}{N} \sum_{n=1}^{N} \Omega_{qni} \cdot \sum_{i \neq n} \sum_{j \neq n} \frac{\sigma_{ij}(n_i)}{\sigma_{ij}} 
  \end{equation}
  
  where, $\sigma_{ij}$ denotes the total shortest paths from node $i$ to node $j$. $\sigma_{ij}(n_i)$ signifies the subset of these paths that traverse through node $n_i$. Meanwhile, $\Omega_{qni}$ characterizes the degree of node $n_i$ in the $q$-th network configuration.

  \item \textit{Algebraic Connectivity:} 
Algebraic connectivity is defined as the second smallest eigenvalue of the normalized Laplacian matrix derived from the network's degree matrix minus its adjacency matrix, as referenced in \cite{ref42}. This metric provides a measure of the network's structural resilience and its ability to withstand faults. Higher values of $\lambda_2$ indicate improved fault tolerance and resilience against network partition, where the network divides into isolated components.

  \item \textit{Percolation Threshold:} The percolation threshold is a critical measure of a system's ability to endure extreme events without power interruptions. It marks a point where the system's behavior shifts, potentially leading to failures. To evaluate the resilience of an electrical distribution system, we calculate the percolation threshold by testing the network's ability to stay connected and deliver power under challenging conditions. By repeatedly removing random edges, we monitor the size of the largest cluster $(S(p))$ and calculate the bond occupation probability $(p)$. More repetitions result in a more reliable estimation of percolation strength \cite{ref43,patwardhan2022machine}:
    \begin{equation}
        P_\infty (p) = \frac{1}{NT} \cdot \sum_{i=1}^{T} S(p)
    \end{equation}
We assess the network's susceptibility through the following calculation:
\begin{equation}
    \chi(p) = \dfrac{\dfrac{1}{N^2 T}\sum\limits_{i=1}^{T} [S(p)]^2-[P_\infty (p)]^2} {P_\infty (p)}
\end{equation}

The percolation threshold $(p_m)$ is the critical p-value at which susceptibility reaches its maximum:
\begin{equation}
    p_m = \arg \{\max_p \chi(p)\}
\end{equation}

In the context of an electrical network, a higher percolation threshold suggests a more resilient network \cite{Dd_resilience}.

  \item \textit{Critical Fraction of Nodes:}  The determination of $f_c$ is of paramount importance from a network analysis perspective. It allows for the assessment of the types of events to which the network could remain resilient and facilitates the optimization of system design to maximize $f_c$. It is computed using the following equation \cite{critical}:
\begin{equation}
f_c = \frac{k}{k_c} - 1
\end{equation}

where, $f_c$ represents the critical fraction of sustainable network damage. $k$ represents the ratio of variance and average degree distribution of the network configuration and $k_c$ is a critical value indicating a threshold for network resilience \cite{critical}.

  \item \textit{Average Path-Length:} 
The average path length, denoted as $l_q$, calculates the minimum number of branches, denoted as $d$, required to travel from one node $n_i$ to another node $n_j$ in the network.\cite{Main_percolation}:
 \begin{equation}
    l_q = \frac{1}{N(N-1)}\sum_{i,j}d(n_i, n_j) 
\end{equation}
This metric provides insight into network reachability and efficiency within the power distribution system.

    \item \textit{Clustering Coefficient:} The clustering coefficient $C_n$ for a specific topology of the graph $G$ is defined as the mean of the local clustering coefficients computed for all the nodes within $G$ \cite{Dd_resilience}:
  \begin{equation}
    C_n = \frac{1}{N}\sum_{i\in N}y_i\left(d_i^2\right) 
    \end{equation}

    where, $y_i$ stands for the count of links connecting neighbors of node $n_i$, and $d_i$ represents the degree of node $n_i$. This value quantifies the likelihood that two neighbors of a node are themselves connected as neighbors.

    \item \textit{Spectral Radius:} The spectral radius $\rho$ is the greatest absolute value among the eigenvalues of the adjacency matrix of $G$ \cite{reference47}:
    \begin{equation}
    \rho = \max_{1\leq i\leq N}|\lambda_i| 
    \end{equation}
    A smaller spectral radius indicates higher network robustness, rendering it more resilient and better defended against cyber-attacks \cite{reference48}.

    \item \textit{Natural Connectivity:} The natural connectivity $\mu$  is a rescaled average eigenvalue of the adjacency matrix of the graph \cite{Main_percolation}:
    
    \begin{equation}
    \mu_n = \ln\left[\frac{1}{n}\sum_{k=1}^{n}e^{\mu_k}\right] 
    \end{equation}
    where $\mu_k$ is the $k$-th eigenvalue of the adjacency matrix. Greater values of $\mu$ correspond to increased strength and robustness against the removal of branches or nodes within the graph.
    \item \textit{Degree Distribution:} The degree of each node is determined by counting the number of neighboring nodes connected to that specific node. The degree distribution of an electrical distribution system serves as an indicator of its topological resilience, reflecting the diversity of connections within the distribution system. The degree distribution for all nodes is expressed as follows:
    \begin{equation}
    k = \dfrac{2 |E|}{|N|} \quad 
    \end{equation}
    \item \textit{Effective Graph Resistance:} It is used to evaluate the connectivity and vulnerability of a network to disruptions or failures and calculated using the formula:
\begin{equation}
R_{\text{eff}}(i, j) = d\frac{2 \cdot S(i, j)}{I}
\end{equation}

where, $R_{\text{eff}}(i, j)$ is the effective resistance between nodes $i$ and $j$. $S(i, j)$ is the sum of effective resistances between $i$ and $j$ through all other nodes in the graph. $I$ is the total effective resistance between $i$ and $j$  when all edges are removed from the graph \cite{egr}.

\end{enumerate}
A substantial amount of research has been dedicated to developing resilience metrics based on the complex network parameters discussed earlier as detailed in Table \ref{tab:complex}. Researchers have employed these metrics by integrating various network features and assigning weights to each feature to assess network resilience.

\begin{table}[h]
\centering
\caption{Complex network parameters used in literature for resilience evaluation.}
\label{tab:complex}
  \renewcommand{\arraystretch}{1.5}
\begin{tabular}{|c|l|}
\hline
\textbf{Reference} & \textbf{Parameters} \\
\hline
\cite{critical} & $D_g$, $ACB$, $\lambda_2$, $f_c$, $l_q$, $C_n$ \\
\hline
\cite{Main_percolation} & $D_g$, $ACB$, $\lambda_2$, $f_c$, $l_q$, $C_n$, $\rho$, $\mu_n$ \\
\hline
\cite{cl4} & $D_g$, $ACB$, $\lambda_2$, $f_c$, $l_q$, $C_n$, $\rho$, $\mu_n$ \\
\hline
\cite{anurag2} & $D_g$, $ACB$, $\lambda_2$, $f_c$, $l_q$, $C_n$ \\
\hline
\cite{Dd_resilience} & $p_m$, $C_n$ \\
\hline
\cite{pc1} & $p_m$, $f_c$ \\
\hline
\cite{ANU1} & $\lambda_2$, $C_n$,$l_q$ \\
\hline
\cite{rocchetta2022enhancing} & $\mu$, $\rho$, $\lambda_2$, $R_{eff}$ \\
\hline

\end{tabular}
\end{table}

\section{Data Driven Applications in
Power System Resilience}
\label{section:data_driven}

\subsection{Data-Driven Methods in Power Systems}
While numerous researchers have previously focused on enhancing resilience through purely graph-based topological approaches, they often overlook the dynamic effects of the system, neglect critical technical parameters, and disregard various other influencing factors \cite{rocchetta2022enhancing}. Most of the resilience analysis and enhancement methods based on physical modeling are challenging due to higher levels of complexity and uncertainty in the system.  Nevertheless, for extreme disasters with intricate mechanisms that could not be precisely described by explicit physical laws, the practical application of these physical methods is constrained \cite{zhang2020optimal}. Modernization and automation allow the system to handle the large sets of data and information which is obtained from the different sensors placed at different levels of the power systems. In recent years, electric power companies have collected an increasing amount of data which has given rise to the use of data-driven techniques for predicting power system faults during typhoon disasters. These datasets give detailed measurements about the grid health and state parameter values with their geographical information constantly to the operator of the system. Additionally, observations of harsh weather conditions, power outages, transient responses, and alarms are included in this data, and they may be helpful for strengthening resilience. Additionally, the data may also contain synchronous information from devices like PMU and other past operating data of the power systems \cite{xie2020review, dd_pmu}. In order to improve the resilience of the system, it is essential to analyze this vast data and discover the hidden value in it which requires the deployment of artificial intelligence (AI) technology-based models like machine learning (ML), deep learning (DL) and reinforcement learning models (RL). These data-based models convert the input data into meaningful output instances by the past data as a process of learning from the known examples. The training of these models with the help of the above-mentioned historical data should be done in offline mode and deployed in the field with real-time input data.

\subsection{Applications of Data-Driven Methods for Power System Resilience}
Rozhin et al. of \cite{eskandarpour2016machine} considered the two states' damage (outage) and operation (in service) predictions based on the weather data obtained from the weather forecasting agencies. Two decision-making factors (hurricane wind speed and the component distance from the hurricane center) are considered for obtaining the cost function for making the decision boundary. Logistic regression is used for the prediction the weather events which is simple and effective but requires a large amount of data as compared to the other models like support vector machines.

\begin{table}[]
\centering
\caption{Static and dynamic variables.}
\label{tab:variable classification}
  \renewcommand{\arraystretch}{1.5}
\begin{tabular}{|l|l|l|}
\hline
\textbf{Variable name} & \textbf{Variable Type} & \textbf{Data Source} \\
\hline
         Rainfall process & Dynamic & Rainstorm waterlogging\\
         \hline
         Terrain type & Static & Public information platform\\
         \hline
         Rainstorm intensity & Static & Electric power company\\
         \hline
         Ponding water depth & Dynamic & Rainstorm Waterlogging\\
         \hline
         Forest coverage rate & Static & Public information platform\\
         \hline
         Number of customers & Static & Electric power company\\
\hline

\end{tabular}
\end{table}

\begin{table*}[]
    \centering
    \caption{Data-driven resilience enhancement and outage predictions.}
    \label{ML table}
    \renewcommand{\arraystretch}{1.5}
    \begin{tabular}{|c|l|l|}
    \hline
    \textbf{Reference}  & \textbf{Proposed Methodology} & \textbf{Key Observations}\\
    \hline
    \cite{cerrai2019predicting} & Outage prediction model based on  &  Outage prediction errors are minimized by implementing an optimization technique.\\
    &  regression tree classifier & The model is validated for 120 storms of different intensities and characteristics.\\
    & & This model suffers from overfitting, high variance and low bias.\\
    \hline
    \cite{liu2007statistical} & Generalized additive models (GAM) \& & Predicted the electricity downtime in case of hurricanes and ice storms.\\
    & Accelerated failure time (AFT) & Tree-trimming covariates are neglected due to the unavailability of the data.\\
    \hline
    \cite{yuan2020development} & Random forest trees & More accurate resolution and high spatial resolution.\\
    & & Accuracy in estimating the number of customers without power \& damaged poles.\\
    & & This model assumes that customer outage and pole damage occur in all grid cells.\\
    \hline
    \cite{kamruzzaman2021deep} & Multi-agent soft actor critic & Scalability issues are improved and verified up to IEEE 300 bus system.\\
    & based deep reinforcement & Higher dimensionality problems are also mitigated due to the interaction of agent with local \\
    & learning & sub-systems.\\ 
    & & The action types provided are both continuous and discrete.\\
    \hline
    \cite{mohammadian2021data} & Bayes decision classifier & Resampling method helps to overcome the imbalance problem.\\
    & with resampling methods & The output may also be used for preventive measures like generation rescheduling.\\
    & & Adaptability for use in diverse geographic areas, relying solely on publicly accessible data.\\
    \hline
    \cite{bahrami2022novel} & Markov decision models & It considers the interaction between the tree and line.\\
    & & Preventive mitigation is employed by constructing the islands prior to the storms.\\
    & & A trade-off always exist between the computational burden and solution accuracy.\\
    & & The performance evaluation metrics are limited to the placement of tree and lines.\\
    \hline
    \cite{spiegel2021hybrid} & Decision tree with &  Tree based approaches delivers feasible solutions.\\
    & hybrid optimization methods & Scheduling problem is solved by optimization methods and enables external grid models.\\
    & & It is assumed that deterministic forecasts are accessible, but are unknown to realize during\\
    & & scheduling.\\
    \hline
    \cite{hussain2019impact} & Deterministic data-driven model & The dynamics of uncertainties are captured using the historical data.\\
    & & The model is verified with Monte Carlo Simulations to ensure survivability of loads.\\
    & & Huge reduction in load shedding with a slight increase in increase in normal operation cost.\\
    \hline
    \cite{abdelmalak2022enhancing} & Predictive recursive Markov process & Integrate spatiotemporal characteristics of wildfires into probabilistic behavior of power grid.\\
    & & Minimizes cost of load shedding and operational costs under specified modeling constraints.\\
    & & Accuracy and effectiveness are validated with IEEE 30-bus system.\\
    \hline
    \end{tabular}
\end{table*}

Recent studies reveal that by utilizing multi-task learning (MTL) and multi-instance learning (MIL) a novel ML methodology \cite{alqudah2023enhancing} for predicting the power outages and also for providing the leading indicators (spatiotemporal precursors) to the grid operators for mitigation in power outages. This approach uses sparse settings for integrating multi-level data, such as local weather, global demand, and forecast data. The grid operators successfully implement their outage mitigation plans since the suggested model accurately predicts incidents up to several hours in advance. But this model neglects a few factors such as equipment status and its degradation that contribute to power outages and also strictly holds to the assumption that weather conditions are the only cause of the outages while man-made errors and machinery failure cases are neglected.

A method proposed in \cite{lei2023evaluation} by merging both particle swarm optimization (PSO) and the least square support vector machines (LSSVM)  for predicting the effects of the economic resilience index. The results show better prediction as compared with the different models such as the backpropagation neural network and the LSSVM individually. The main advantages of this model are the removal of redundant information from the index data, and the fitting degree was also significantly improved. Dependency on the human experience is the main concern of adopting PSO \cite{K2024109502}.

The prediction model for the disconnected lines due to the heavy rain storms is proposed in \cite{lei2023evaluation}, establishing the rain storm waterlogging model by using the unimodal rainfall process and the rainfall intensity. A knowledge data-driven model is used with the help of dual path prediction compromising three parts namely, stochastic feature extraction (using feed-forward neural network and leaky ReLU activation function), dynamic feature extraction (uses the recurrent neural networks such as gated recurrent Units), and the feature fusion process where multi-head attention models are utilized. This fusion is deployed based on scaled dot product fusion for analyzing the potential different relationships between the various indices of data from diverse points of view. The classification of different variables and their source locations are discussed in Table \ref{tab:variable classification}.

An ML-based method for detecting and interpreting the faults in vulnerable grids is used in \cite{eikeland2021detecting}. The primary objective of this project centers around investigating an Arctic Norwegian community that has encountered a significant number of faults due to unexplained issues. Since the classifiers work well for this type of problem, but providing the set of variables for explaining the individual fault is also important where the classifiers failed. This issue was resolved by adopting a technique to interpret the decision-making process of a DL model called integrated gradients. The proposed approach provides an opportunity to acquire in-depth insights into the specific fault's occurrence. These insights are invaluable for distribution system operators as they could use them to develop strategies aimed at preventing and minimizing power disruptions. Some more recent applications of the data-based methods in resilience concept are provided in \ref{ML table}.

\subsection{Predictive Risk Analytics for Resilience Assessment}

In response to the numerous definitions and mitigation proposed in the literature, two significant developments are offered \cite{leite2019resiliency}. One associates with the ways of improving the operator situational awareness by analyzing the impacts of the actions taken by the operator for achieving improved grid resilience through mitigation strategies prioritization. The latter is improving the controllability, observability, and flexibility of the operation of the grid by preventive measures, especially in the case of HILP events. The combination of these two developments leads to improved resilience in two ways: 
\begin{enumerate}
    \item Usage of proper graphic tools, visualization contours, animated arrows, dynamic size charts, etc.,
    \item Another component involves a targeted decision-making tool that provides a predictive evaluation of the evolving conditions during severe weather events, enabling the formulation of preventive mitigation strategies to minimize risks.
\end{enumerate}

Predictive risk analytics encompass a range of metric parameters forecasted using diverse AI techniques. They also unveil the power system's equipment vulnerabilities and its responsiveness to weather fluctuations.  A comprehensive breakdown of the risk analytics and the AI models utilizing these analytics is presented in Table \ref{predictive analytics}.

\begin{table}[]
    \centering
    \caption{Data-driven predictive risk analytics utilizing advanced models.}
    \renewcommand{\arraystretch}{1.5}
    \label{predictive analytics}
    \begin{tabular}{|l|l|c|}
    \hline
    \textbf{Predictive Risk Analytics} & \textbf{Type of model} & \textbf{Reference}\\
    \hline
     Weather-based risk assessment & Naive bayes model & \cite{leite2019resiliency}\\
     \hline
     High-risk weather threats & Probabilistic forecasts & \cite{dehghanian2018predictive}\\
     \hline
     Line fault detection & CNN and RNN models & \cite{almasoudi2023enhancing}\\
     \hline
     Grid vulnerabilities to hazards & Reinforcement learning & \cite{bagheri2023advancements}\\
     \hline
     Short-term electricity prices & Neural network approach & \cite{catalao2007short}\\
     \hline
     Power outage estimation & Support vector machine & \cite{eskandarpour2017improving}\\
     \hline
     Microgrid generation  & Sensor data-driven model & \cite{fallahi2021predictive}\\
     \& maintenance & &\\
     \hline
    \end{tabular}
\end{table}

\section{Resilience Enhancement Techniques}
\label{section:enhancement}
\begin{figure}
  \centering
  \includegraphics[width=3.4in]{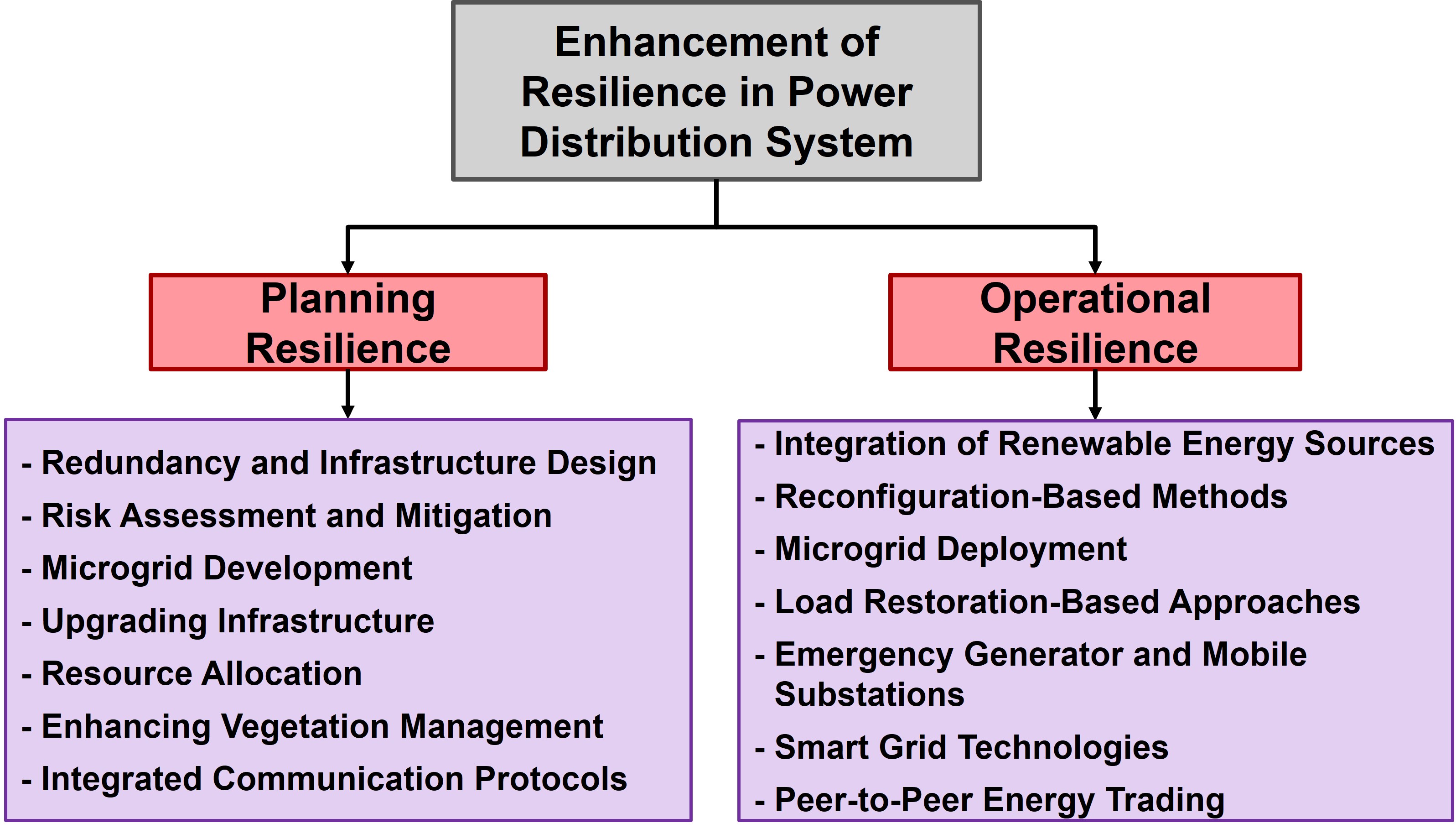}
  \caption{Resilience enhancement techniques.}
  \label{fig:resilience_enhancement}
\end{figure}

Resilience enhancement techniques are employed to improve the ability of a system or network to withstand and recover from various disruptions or adversities. These techniques aim to increase the system's capability to adapt, absorb shocks, and continue functioning under adverse conditions. Resilience enhancement techniques are implemented through two key phases: the planning phase and the operational phase,  as shown in Fig. \ref{fig:resilience_enhancement}.

\subsection{Planning Approaches for Resilience Enhancement}

Planning-based resilience enhancement methods play a crucial role in power distribution systems against disruptions and challenges. These proactive methods cover a broad spectrum of measures, including undergrounding distribution and transmission lines, building redundant routes for power distribution, and upgrading critical infrastructure components with more robust materials  \cite{hard1, hard2, hard3, hard4, hard5}. By investing in elevating substations, incorporating backup generators, and installing remote control switches, power utilities enhance their ability to maintain operational continuity during adverse conditions, such as severe weather events or flooding \cite{design1,design2, design3,design4,design5,design6}. Effective vegetation management further reduces outages caused by falling trees and branches \cite{veg1,veg3,veg2,veg4}, while optimizing the placement and sizing of battery energy storage units and renewable energy sources contributes to grid sustainability. Secure communication protocols ensure that critical data flows securely between grid components, guarding against cyber threats. These planning-based resilience enhancement measures are pivotal in preparing power distribution systems to endure disruptions, providing a reliable and uninterrupted electricity supply to consumers during disruptions. The latest papers for the years 2022-2024 under various categories for planning strategies are tabulated in Table \ref{tab:planning_strategies}.

\subsection{Operation Based Resilience Enhancement Technique}

Operational resilience strategies in power distribution systems are vital to minimize the impact of adverse events and disruptions. They offer dynamic solutions to reduce downtime and ensure a dependable supply of electricity during challenging circumstances. These strategies encompass various approaches, including network reconfiguration \cite{emr1,emr2,emr3,emr4,emr5,emr6,emr7,emr8}, microgrid formation \cite{mg1,mg3, nmg3, NMG1, NMG2, omg1, omg2, omg3, omg4}, the utilization of mobile emergency resources and energy storage units \cite{mobile1,mobile2,ms1,ms2,ms3,ms4,ms5,ms6}, integration of renewable energy sources \cite{opr10,opr11,opr12,opr3,opr4,opr5,opr6,opr7,opr8,opr9}, and peer-to-peer energy trading \cite{p2p1,p2p2,p2p3,p2p4,p2p5,p2p6,p2pnew1,p2pnew2} as tabulated in Table \ref{tab:operational_strategies}. Network reconfiguration methods optimize the layout of the grid, while microgrid formation divides the system into smaller, self-reliable units capable of operating independently or in conjunction with the main grid. The mobilization of mobile energy resources and energy storage units ensures swift response during normal and emergency conditions. Load restoration strategies expedite the process of restoring power post-disruption, aided by distributed generation and proactive repair crew dispatch. Moreover, situational awareness-based frameworks enhance resilience by predicting and responding to power outages.  P2P energy trading is integrated with microgrids, which are smaller and localized energy distribution systems that operate independently from the main grid during disruptions. This enhances the resilience of the community's energy infrastructure. By allowing individuals and communities to generate income from surplus energy, P2P energy trading could enhance economic resilience. It creates a potential source of revenue and could reduce energy costs for participants. These strategies collectively enhance the reliability of power distribution systems, safeguarding critical infrastructure and ensuring uninterrupted power supply during disruptions.

\begin{table}[h!]
  \centering
  \caption{Planning strategies for enhancement of resilience for the years 2022-2024.}
  \label{tab:planning_strategies}
  \renewcommand{\arraystretch}{1.5}
  \begin{tabular}
  {|p{4.5cm}|p{1.5cm}|}
    \hline
    \textbf{Techniques} & \textbf{References}\\
    \hline
    Redundancy and Infrastructure Design & \cite{hard1, hard2, hard3, hard4, hard5, design1,design2, design3,design4,design5,design6} \\
    \hline
    Risk Assessment and Mitigation & \cite{risk1,risk2,risk3,risk4,risk5}\\
    \hline
    Microgrid Development  & \cite{mgnew1,mgnew2,mgnew3,mgnew4,mgnew5,mgnew6,mgnew7,mgnew8}\\
    \hline
    Resource Allocation & \cite{ra1,ra2,ra3,ra4,ra5,ra6, ra7} \\
    \hline
    Enhancing Vegetation Management & \cite{veg1,veg3,veg2,veg4} \\
    \hline
    Integrated Communication Protocols & \cite{com2,com1,com4,com5,com6}\\

    \hline
    
\end{tabular}
\end{table}

\begin{table}[h!]
  \centering
  \caption{Operational strategies for enhancement of resilience for the years 2022-2024.}
 \renewcommand{\arraystretch}{2} \label{tab:operational_strategies}
  \begin{tabular}
  {|p{5cm}|p{3cm}|}
    \hline
    \textbf{Techniques} & \textbf{References}\\
    \hline
    Integration of Renewable Energy Sources & \cite{opr10,opr11,opr12,opr2,opr3,opr4,opr1,opr5,opr6,opr7,opr8,opr9} \\
    \hline
    Reconfiguration-Based Methods & \cite{emr1,emr2,emr3,emr4,emr5,emr6,emr7,emr8}\\
    \hline
    Microgrid Deployment & \cite{mg1,mg3, nmg3, NMG1, NMG2, omg1, omg2, omg3, omg4} \\
    \hline
    Load Restoration- Based Approaches & \cite{lr1,lr2,lr3,lr4,lr5,lr6,lr7}\\
    \hline
    Mobile Emergency Generators & \cite{mobile1,mobile2,ms2,ms3,ms4,ms5,ms6} \\
    \hline
    Peer-to-peer Energy Trading & \cite{p2p1,p2p2,p2p3,p2p4,p2p5,p2p6,p2pnew1,p2pnew2,resiliencycost} \\
    \hline
\end{tabular}
\end{table}

\begin{table*}[h!]
  \centering
  \caption{A comprehensive summary of the literature review focusing on resilience-oriented microgrid scheduling and energy management.}
  \label{tab:microgrid_strategies}
  \renewcommand{\arraystretch}{1.5}
  \begin{tabular}
  {|p{2.8cm}|p{4cm}|p{6.5cm}|c|}
    \hline
    \textbf{Features} & \textbf{Methods} & \textbf{Contribution} & \textbf{Reference} \\
    \hline
    Microgrid Formation Strategies &  Backtracking Search Optimization & - A dynamic framework for MG formation that aims to optimize both the probability of successful islanding of MGs and their self-sufficiency & \cite{osama}\\

     & Mixed Integer Programming (MIP) & - Examine the effectiveness and precision of three distinct deterministic MIP formulations in addressing intentional controlled islanding problems with a focus on resilience & \cite{Patsakis}\\ 

     & Stochastic Programming and Information Gap Decision Theory &  - Integrate the coordination of hydrogen energy with renewables and hydrogen transit. An equivalent linear reformulation of the transportation network (TN)-free hydrogen transit model, streamlining the integration of hydrogen transit into the MG formation problem & \cite{mg3}\\

     & Darts Game Theory Optimization Algorithm & - Reduces overall losses, minimizes load shedding, and controls the total cost of restoration while adhering to a range of topological and electrical constraints & \cite{dart}\\

     & Graph Theory and Coalitional Game Theory & - Using weather forecasting and monitoring data, the proposed method identifies optimal placement locations for MERs to ensure rapid response in the aftermath of extreme events. & \cite{GAUTAM2023101095}\\

     & Deep Reinforcement Learning & - A dynamic microgrid formation-based service restoration approach using deep reinforcement learning as a Markov decision process (MDP) is proposed. It addresses operational and structural constraints, with a deep Q-network to derive optimal microgrid formation strategies & \cite{new2}\\
     \hline
     Dealing with Variability and Uncertainty & Alternating Direction Method of Multipliers & - To recognize the inherent uncertainties associated with both energy sources and consumption patterns & \cite{uncertainity1} \\
      & Empirical Mode Decomposition with Long Short-term Memory & - To capture both the long-term capability trends and the associated uncertainty arising from  regenerations directly and simultaneously  & \cite{uncertainity2}\\
       & Bayesian Deep Reinforcement Learning & - To develop resilient control solutions for multi-energy micro-grids, addressing challenges related to managing renewable energy uncertainties, implementing rapid-response control during extreme events, and ensuring efficient performance with limited extreme event data  & \cite{uncertainity3}\\   
     \hline
     Mobile Energy (Emergency) Resources & Hierarchical Multi-agent Reinforcement Learning Method & - To develop a decentralized framework for coordinating the deployment of Mobile Power Sources (MPSs) and Repair Crews (RCs), recognizing the potential breakdown of communication infrastructure during extreme events, and enhancing the overall resilience of microgrid systems & \cite{mobile2}\\
     & A three-stage Distributed Control Method Featuring Rolling Optimization & - To employ optimization techniques to capture time-coupled flexibility, uncertainties, and mobility of MESSs, while addressing detailed network structures and technical constraints. 
 & \cite{mobile1}\\
    & Multi-stage and Spatio-temporal Operation Model & -  To optimize the operation, ensuring that the MBS units are used efficiently to maximize the utilization of renewable resources and minimize operational costs & \cite{mobile2}\\
     \hline
     Networked Microgrids Formation & Multi-agent Reinforcement Learning (MARL) & - A decentralized framework for resilience-oriented coordination of NMGs is introduced. It uses a MARL method that aims to develop an efficient credit assignment scheme for NMGs & \cite{NMG1} \\
     & Optimal Scheduling Model & - To minimize day-ahead costs while considering operational constraints, with a specific emphasis on maintaining sufficient online capacity for all MGs during emergency islanding scenarios caused by events like extreme weather conditions & \cite{NMG2} \\
     & Multi-period Two-stage Scenario-based Stochastic Mixed-integer Linear Programming
 & - To manage the operation of multi-microgrids in a coordinated manner under both emergency and normal conditions, taking into account the impact of each situation on the other & \cite{nmg3} \\
     \hline
         
  \end{tabular}
\end{table*}

\subsubsection{Enhancement Through Microgrids}

\begin{figure}
  \centering
  \includegraphics[width=3.1in]{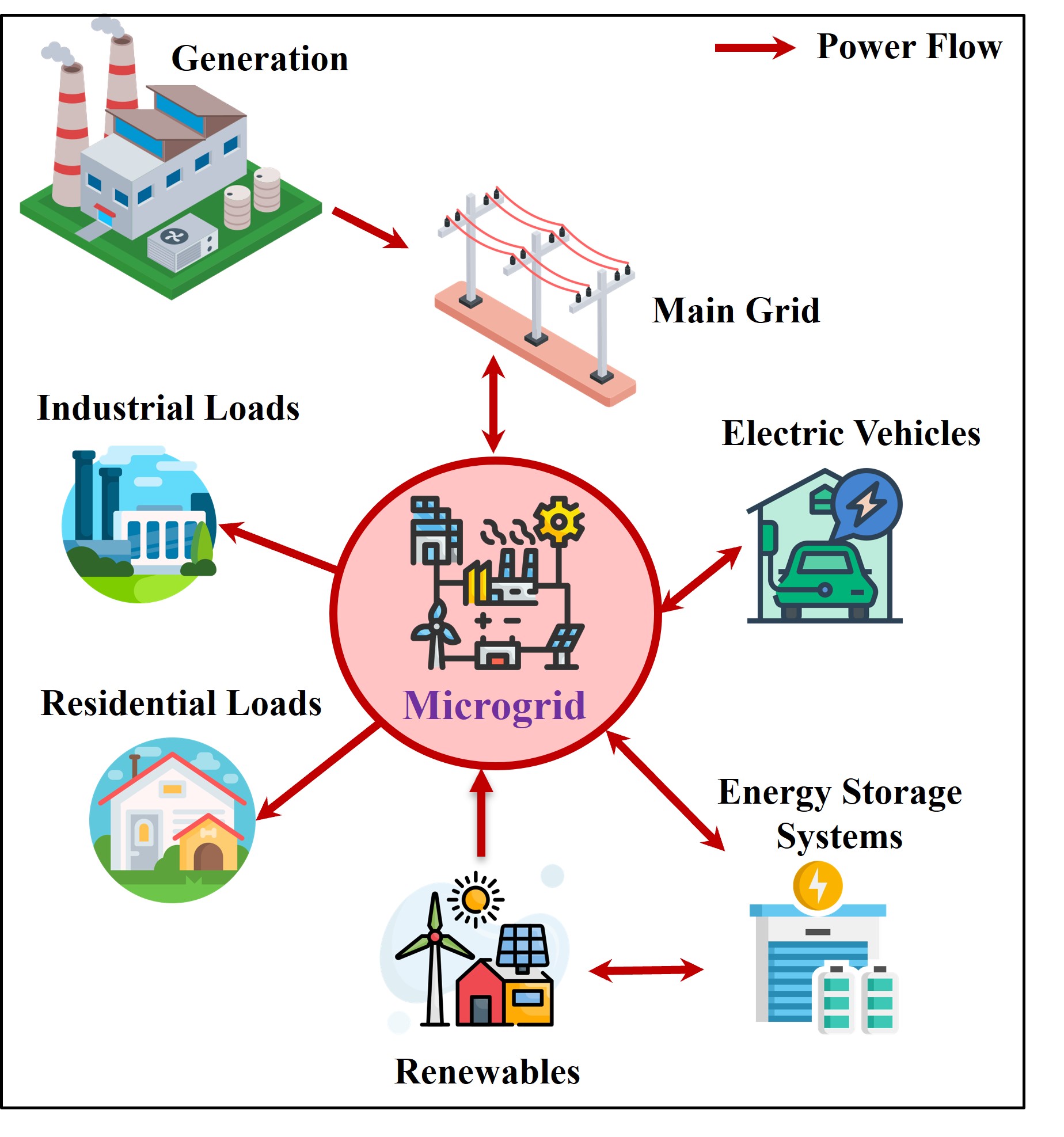}
  \caption{General structure of microgrid.}
  \label{fig:Microgrid}
\end{figure}

The term ``Microgrid" was first introduced by Lasseter and Paigi in their work \cite{mg1}. A microgrid usually consists of a collection of small-scale energy sources and consumers that function as a single, controllable system. The typical structure of a microgrid is shown in Fig. \ref{fig:Microgrid}. The U.S. Department of Energy provides an official definition for microgrids: ``A microgrid is a network of interconnected power sources and distributed energy resources (DERs) that operate within a defined electrical boundary. It behaves like a unified and controllable entity to the larger electrical grid. An important feature of microgrids is their ability to connect to and disconnect from the main grid, allowing them to function either in sync with the grid or in an isolated `island mode' \cite{DOE2023}."

The implementation of the smart grid, along with the growing integration of DERs and remotely controllable switch devices, has led to intentional and controlled islanding from RESs and dispatchable DGs which are considered to be promising strategies for enhancing the resilience of the distribution system against significant events. This design empowers microgrids to supply energy to their local consumers without the need for expensive long-distance transmission infrastructure. Moreover, a microgrid isolates itself from the main grid and continues to operate autonomously, even if it becomes disconnected from the larger power network.

The core concept behind the restoration of electric service via optimal MG formation is the efficient allocation of resources to maximize the total weighted sum of restored loads by locally providing power through DERs. This approach is put into action until the main grid is entirely restored. To execute this strategy effectively, an optimal switching plan is essential. In cases where controllable switches are lacking, the problem becomes equivalent to optimizing the placement of additional switches. Following a major event provided that the communication and monitoring system remains operational, it becomes possible to identify network configuration and system failures. The use of controllable switches to isolate the section of the network with the faulted zone, the self-healing system restructures the network into an optimal configuration of self-sufficient MGs. These MGs could independently supply power to local loads. However, several challenges persist in effectively establishing MGs using RESs and DGs, especially in scenarios involving infrastructure or communication system breakdowns \cite{651}. A detailed enhancement technique proposed by researchers is tabulated in Table \ref{tab:microgrid_strategies}, which highlights the techniques for the formation of microgrids and the methodologies used.

\subsubsection{Enhancement through Renewable Energy Integration}

Enhancing energy systems through renewable energy integration is a critical strategy for achieving sustainable and resilient energy supplies while mitigating the impacts of climate change. This integration includes the integration of renewable energy sources, such as solar, wind, hydro, and geothermal power, into the existing energy infrastructure. Greece faced a wildfire challenge in August 2021 following an extreme heatwave \cite{renewable1}. These wildfires affected 125,000 hectares of forest and cultivable land, along with numerous homes. The number of fires exceeded the 12-year average by 26\%, and the burned area was 450\% larger than usual. The distribution network suffered extensive damage in various regions. For instance, in Evia, 7 medium voltage (MV) lines supplying electricity to 13,000 consumers were damaged. In Attica, 9 MV lines serving 38,000 consumers were affected. To address the electricity supply needs during these disasters, 12 mobile generators having a combined capacity of 2.12 MW were deployed to provide the required power before the network could be fully restored \cite{renewable1}.

There has been extensive work done on providing strategies to integrate DERs. Shiva Poudel et al. proposed approach outlines a resilient restoration strategy that seeks to simultaneously maximize the number of critical loads that could be restored and optimize the restoration time by providing available DERs within the grid \cite{renewable2}. To address this complex problem, a MILP is formulated to maximize the resilience against post-restoration failures. This ensures compliance with critical infrastructure and operational conditions within the grid, all the while maintaining a radial operation for a given open-loop feeder configuration. Mohammad Amin Gilani et al. also used a mixed-integer linear program designed to efficiently restore prioritized loads while adhering to essential topology and operational constraints within the distribution system \cite{renewable3}. The model delves into the dynamic formation of microgrids and the optimal placement of a range of smart grid technologies, including distributed generation sources, demand response programs, wind turbines, and energy storage units. Furthermore, the substantial influence of renewable energy resources, varying load patterns, and the inherent uncertainty associated with these factors on the overall resilience of the distribution system is explored. By considering these complex interactions and employing mathematical optimization, the research aims to enhance the understanding and management of distribution system resilience in the context of evolving smart grid technologies and the integration of renewable energy sources. 

An integrated framework for active distribution systems, focusing on probabilistic models to address extreme events is proposed \cite{renewable4}. By combining probabilistic extreme event modeling, impact assessment, and optimal restoration models within a non-sequential Monte Carlo Simulation framework, the system's resilience against dynamic challenges is comprehensively evaluated. The framework also accounts for the intricate relationships between variables like demand fluctuations, renewable energy generation, and energy storage characteristics. Metrics are proposed to gauge system resilience and quantify outage impacts at specific load points. These metrics and their associated probability distributions offer a valuable basis for informed investment planning, helping decision-makers select the most effective measures to enhance the robustness and reliability of active distribution systems when confronting unforeseen extreme events.

\begin{figure}
  \centering
  \includegraphics[width=3.4in]{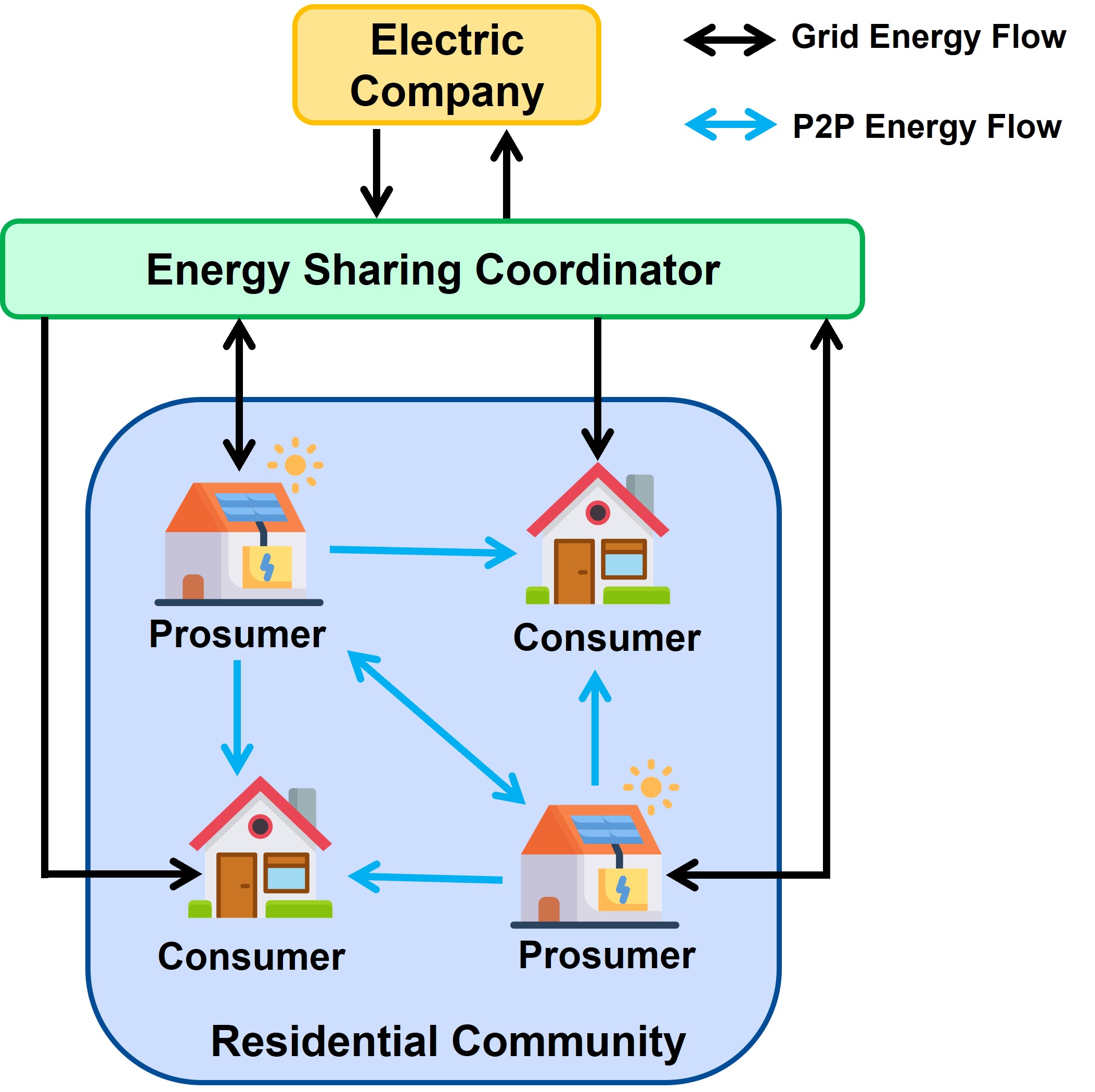}
  \caption{General structure of P2P energy trading model.}
  \label{fig:p2p}
\end{figure}

\subsubsection{Enhancement Through Peer-to-Peer Energy Trading}
Peer-to-peer energy trading is an innovative approach to the energy sector that allows individuals and businesses to buy and sell excess electricity directly to one another. Recently, it has been adopted to enhance the resilience of electrical distribution systems as shown in Fig. \ref{fig:p2p}. A collaborative approach for enhancing the resilience of microgrids through peer-to-peer energy exchanges has been introduced \cite{p2pnew1}. In this advanced model, microgrids function in a networked and cooperative fashion, facilitating the exchange of energy at the most cost-effective rates among themselves. By implementing this innovative model, microgrids could enhance their resilience through peer-to-peer energy transactions while simultaneously reducing their operational expenses, as opposed to operating independently in a non-cooperative manner. The collaborative advantages for both the system operator and end-users were put forward \cite{p2p1} focusing on a) realizing economic benefits through active market participation,
b) strengthening system resilience, c) extending the lifespan of batteries, and d) reducing carbon emissions. This approach optimizes battery sizing and operation while classifying and prioritizing end-users. The proposed method has enhanced the system resilience by up to 80\% and extended battery life by 32\% to 37\%.

A distributed peer-to-peer energy management strategy that is resilient to Byzantine faults has been designed \cite{p2pnew2}. The primary objective of this approach is to minimize the total cost incurred by prosumers while taking into account a range of constraints, including limits on power output, power line congestion, and voltage magnitude. A transactive multiple MGs (TMMGs) concept is proposed in \cite{p2p2}. It introduces an innovative framework for resilient scheduling of TMMGs, emphasizing the need for reliable operation during long-duration disconnections from the main grid. To optimize the day-ahead scheduling of TMMGs, the study employs a cooperative game approach, taking into account the presence of multiple MGs within the TMMG and their potential for energy transactions through a local market. This local energy market is a pivotal component, allowing the TMMG operator to collect bid-offer amounts from individual MGs and establish hourly transaction prices. The proposed game aims to minimize the operation cost of MGs during normal operation mode (NOM) and ensure that the TMMG effectively serves its load during resiliency operation mode (ROM). The equilibrium point sought in this research ensures that all participating MGs adopt acceptable strategies while adhering to the constraints of both NOM and ROM, thereby ensuring efficient and reliable operation. Another framework for enhancing the resilience of networked microgrids is proposed, which is implemented in two sequential stages \cite{p2pnew3}. Firstly, each individual microgrid seeks to maximize its exported power, contributing to the overall resilience of NMGs. Importantly, this optimization is carried out while adhering to an individual rationality constraint, ensuring that the operational costs for MGs engaging in P2P energy sharing do not surpass those incurred when they operate independently. In the second stage, a decentralized approach is employed, leveraging an average consensus algorithm and a proportional sharing assumption. This enables each MG to determine its imported and exported power levels. Numerical results from the study showcase the effectiveness of this framework in bolstering the overall resilience of NMGs. Furthermore, by maintaining the individual rationality constraint, the results reveal that MGs tend to favor the P2P networked operation mode over the individual operation, highlighting the advantages of cooperative energy sharing in terms of cost-effectiveness and resilience.

Thereby, P2P energy trading holds the potential to reshape the energy sector by empowering individuals and communities to take a more active role in energy generation and consumption. However, its success depends on the willingness of governments, utilities, and consumers to adapt to this emerging energy interaction.

\section{RESEARCH GAPS, CHALLENGES, AND
FUTURE DIRECTIONS}
\label{section:future}

\subsection{Challenges and Gaps}

Upon reviewing the literature, several challenges and gaps in the existing literature become apparent, which necessitates further exploration and discussion in the future:

\begin{enumerate}[label=\arabic*.]
    
    \item \textbf{Lack of Universal Resilience Definition}: While resilience is commonly associated with attributes like absorptivity, adaptability, and recoverability, there is no universally accepted definition or standard for resilience. This lack of consensus poses a challenge for researchers and practitioners.

    \item \textbf{Shortage of Comprehensive Resilience Indicators}: The available resilience indicators predominantly focus on network absorption and restoration capability. To grasp the true nature of power systems, there is a need for indicators that also encompass predictive and adaptive capabilities, encompassing aspects like system characteristics, quantitative and qualitative analysis, and deterministic and probabilistic considerations.

    \item \textbf{Challenges in Data Accuracy}: Weather-related forecasting methods exhibit low accuracy, and reliance on historical data from a single area assumes complete data reliability. To address this issue, researchers must account for calibration, communication errors, uncertainties, and the need for developing cyber-attack simulation methods due to data limitations and ambiguities.

    \item \textbf{Incomplete Failure Modeling}: Resilience assessments often rely on fragility curves to model failure. However, these curves tend to overlook side effects of events, necessitating further research in this domain.

    \item \textbf{Inadequate Modeling of Resilience Improvement Strategies}: There remains an insufficiency in the study of modeling for resilience improvement strategies. This area of research could be expanded to enhance our understanding.

    \item \textbf{Neglected Preventive Operational Strategies}: Current resilience improvement methods tend to overlook preventive operational strategies. Integrating such strategies significantly reduces damage and improves load management during events.

    \item \textbf{Insufficient Focus on Cost-related Issues}: The financial aspects of resilience, particularly the cost-effectiveness of emergency resources in relation to electricity cost, warrant more extensive investigation.

    \item \textbf{Relationship Among Occurrences}: It is crucial to ensure that an improvement strategy designed for one event does not compromise the network's capability during another event. Understanding the interplay between different events is essential.

    \item \textbf{Critical Loads in Isolated Areas}: The power supply of critical loads in areas lacking conventional power sources, where mobile energy sources are required, has been largely overlooked in resilience modeling and improvement techniques.

    \item \textbf{Complex Interdependencies in Electric Systems}: Electric sources are inherently intertwined with other utilities such as water, gas, and communications. The study of resilience improvement methods in these interdependent systems is in its infancy due to the complexities involved.

    \item \textbf{Underutilized Frameworks}: None of the proposed metrics align with the proposed frameworks, leading to a lack of generalizability and practical real-world application.

\end{enumerate}

These challenges present a roadmap for future research and underscore the need for a comprehensive and multifaceted approach to address the resilience of electrical energy systems in a rapidly evolving and interconnected world.

\begin{enumerate}[label=\arabic*)]
    \item \textbf{Interdependence of Critical Infrastructures:} Traditional approaches to power system resilience have treated it as an isolated entity. However, it is imperative to acknowledge that the performance of the power system is intricately linked with other critical infrastructures, such as telecommunications, transportation, water, oil, and natural gas systems. Future research should focus on assessing the interdependencies among these systems and adopting a holistic approach to simultaneously manage the resilience of all infrastructures.

    \item \textbf{Comprehensive Resilience Metrics:} The development of comprehensive metrics and indices for evaluating power system resilience is essential. These metrics should consider various dimensions of resilience, encompassing factors like reliability, adaptability, and recovery capabilities.

    \item \textbf{Cost-Effectiveness Analysis:} Investigating the cost-effectiveness of strategies to enhance power system resilience in the face of severe HILP events is crucial. This involves identifying models that effectively incorporate economic factors into resilience enhancement solutions.

    \item \textbf{Holistic Resilience Models:} Power system operators require flexible models that allow them to optimally allocate and utilize resilience resources during different events. These models should consider various strategies and resources available to enhance resilience.

    \item \textbf{Renewable Energy Integration:} Given the increasing integration of renewable energy sources into power systems, future research should focus on developing strategies specifically designed for power systems with a substantial share of RESs. These strategies should address unique challenges associated with renewables.

    \item \textbf{Long-Term Planning with DERs:} DERs play a significant role in modern power systems. Investigating how long-term planning, considering factors like the state of health of energy storage systems and electric vehicle batteries, could impact power system resilience is vital.

    \item \textbf{Cyber Resilience:} As information and communications technologies become more integral to power systems, the importance of cyber resilience grows. Research should focus on methods and approaches that consider cybersecurity issues to ensure the stability and security of the power system.

    \item \textbf{Data-Driven Resilience Solutions:} Utilizing accurate historical data, possibly at an hourly resolution, related to the intensities of various events in specific geographic areas helps in selecting the most suitable and least-risk resilience improvement solutions.

    \item \textbf{Climate Models and Geospatial Analysis:} To identify the most likely HILP events in a particular region, researchers consider climate models and conduct geospatial analyses. This approach provides valuable insights into the geographic-specific threats that a power system may face, enabling proactive resilience planning.
\end{enumerate}

\section{Conclusion}
\label{section:Conclusion}

In this comprehensive review, we have discussed the dynamic and critical areas of power system resilience, emphasizing its significance in an era of increasing challenges and disruptions. The comparison of reliability with resilience has presented differences by highlighting the need for a paradigm shift in power system planning and operation to account for the unexpected and severe disturbances that modern grids face. We explored the conceptual foundations of power system resilience, providing clear definitions and using the key characteristics that set it apart from traditional reliability. Resilience is not merely about minimizing downtime; it's about adaptability, recovery, and maintaining critical services even in the face of adversity. The review has discussed two significant frameworks for assessing power system resilience: one developed by Sandia National Laboratory and the other by RAND Corporation. These frameworks offer a structured approach to defining and quantifying resilience metrics, enabling a more systematic assessment of power system resilience.

Quantitative metrics for power system resilience were discussed, including energy or demand served, loss of load metrics, value at risk metrics, and others. These metrics provide the means to measure, compare, and improve power system resilience quantitatively, a vital step in making resilience a practical goal for power system operators and planners. Complex network theory's application in power system resilience provides an understanding of the interconnected nature of the grid and how disruptions propagate. This knowledge is crucial in designing resilient power systems capable of containing and mitigating disturbances.

Furthermore, the integration of data-driven methods and machine learning into the power system resilience domain has immense potential. Case studies showcased the ability of these technologies to predict and assess risks, enhance situational awareness, and expedite response and recovery efforts. The review also explored various techniques for enhancing power system resilience. Planning approaches, operation-based strategies, microgrids, renewable energy integration, and peer-to-peer energy trading were all examined as methods to fortify power systems against disruptions. As we conclude, it's evident that the field of power system resilience is evolving rapidly. However, several challenges and research gaps persist, demanding further attention and investigation. These include developing standardized resilience metrics, accounting for evolving grid technologies, and addressing the human and institutional factors that play a crucial role in resilience. This review paper serves as a foundational resource for researchers, policymakers, and industry professionals. It offers valuable insights into resilient power systems and provides a roadmap for building more robust and adaptive grids capable of withstanding and recovering from unforeseen disruptions. Thus, power system resilience is a necessity for ensuring the reliability and security of our energy infrastructure in an ever-changing world.

\bibliographystyle{IEEEtran}
\bibliography{main}

\end{document}